\documentclass[submission,copyright,creativecommons]{eptcs}
\providecommand{\event}{EXPRESS/SOS 2018}

\usepackage{amsmath}

\usepackage{amsfonts}
\usepackage{amssymb}
\usepackage{amsthm}
\usepackage{verbatim} 
\usepackage{stmaryrd}
\usepackage{paralist}
\usepackage{relsize}
\usepackage{cite}
\usepackage{appendix}
\usepackage{url}
\usepackage[]{units}
\usepackage{tikz}
\usetikzlibrary{calc,arrows,positioning}
\usepackage{array}
\usepackage{color}
\usepackage{empheq}
\usepackage{tabularx}
\usepackage{multicol}
\usepackage{caption}
\usepackage{wasysym}
\allowdisplaybreaks

\newif\ifdraft\drafttrue

\DeclareFontFamily{U}{mathb}{\hyphenchar\font45}
\DeclareFontShape{U}{mathb}{m}{n}{
      <5> <6> <7> <8> <9> <10> gen * mathb
      <10.95> mathb10 <12> <14.4> <17.28> <20.74> <24.88> mathb12
}{}
\DeclareSymbolFont{mathb}{U}{mathb}{m}{n}
\DeclareMathSymbol{\sqdoublecup} {2}{mathb}{"5F} % From mathabx.dcl
\DeclareMathSymbol{\boxplus} {2}{mathb}{"60} % mathabx.dcl

\newcommand{\trans}[1][]{\xrightarrow{\, {#1} \, }}
\newcommand{\ntrans}[1][]{\mathrel{{\trans[#1]}\makebox[0em][r]{$\not$\hspace{2ex}}}{\!}}

\newcommand{\support}{\mathsf{supp}}

\newcommand{\Act}{\mathcal A}

\newcommand{\proc}{\mathbf{S}}
\newcommand{\ProbDist}[1]{\Delta(#1)}

\newcommand{\nil}{\mathrm{nil}}

\newcommand{\Z}{\mathcal Z}
\newcommand{\corr}[1]{\mathrm{corr}_{#1}}

\newcommand{\ctrans}[1][]{\stackrel{#1}{\mathlarger{\twoheadrightarrow}}}

\newcommand{\pr}{\mathrm{Pr}}

\newcommand{\Tr}{\sim_{\mathrm{Tr}}}

\newcommand{\TrN}{\Tr^{\mathbf{N}}}
\newcommand{\TrP}{\Tr^{\mathbf{P}}}

\newcommand{\res}{\mathrm{Res}^{\mathrm{det}}}

\newcommand{\resx}{\mathrm{Res}^{\mathrm{x}}}

\newcommand{\trdist}{\sim_{\mathrm{Tr,dis}}^{\mathrm{det}}}

\newcommand{\trdistct}{\sim_{\mathrm{Tr,dis}}^{\mathrm{rand}}}
\newcommand{\sqdistx}{\sqsubseteq_{\mathrm{Tr,dis}}^{\mathrm{x}}}
\newcommand{\trdistx}{\sim_{\mathrm{Tr,dis}}^{\mathrm{x}}}

\newcommand{\trdistdist}{\mathbf{m}_{\mathrm{Tr,dis}}^{\lambda,\mathrm{det}}}
\newcommand{\trdistpredist}{\mathbf{h}_{\mathrm{Tr,dis}}^{\lambda,\mathrm{det}}}
\newcommand{\trdistdistct}{\mathbf{m}_{\mathrm{Tr,dis}}^{\lambda,\mathrm{rand}}}
\newcommand{\trdistpredistct}{\mathbf{h}_{\mathrm{Tr,dis}}^{\lambda,\mathrm{rand}}}
\newcommand{\trdistdistx}{\mathbf{m}_{\mathrm{Tr,dis}}^{\lambda,\mathrm{x}}}
\newcommand{\trdistpredistx}{\mathbf{h}_{\mathrm{Tr,dis}}^{\lambda,\mathrm{x}}}

\newcommand{\tbt}{\sim_{\mathrm{Tr,tbt}}^{\mathrm{det}}}
\newcommand{\sqtbtct}{\sqsubseteq_{\mathrm{Tr,tbt}}^{\mathrm{rand}}}
\newcommand{\tbtct}{\sim_{\mathrm{Tr,tbt}}^{\mathrm{rand}}}
\newcommand{\sqtbtx}{\sqsubseteq_{\mathrm{Tr,tbt}}^{x}}
\newcommand{\tbtx}{\sim_{\mathrm{Tr,tbt}}^{x}}

\newcommand{\tbtdist}{\mathbf{m}_{\mathrm{Tr,tbt}}^{\lambda,\mathrm{det}}}
\newcommand{\tbtpredist}[1]{\mathbf{h}_{\mathrm{Tr,tbt}}^{{#1}\lambda,\mathrm{det}}}
\newcommand{\tbtdistct}{\mathbf{m}_{\mathrm{Tr,tbt}}^{\lambda,\mathrm{rand}}}
\newcommand{\tbtpredistct}[1]{\mathbf{h}_{\mathrm{Tr,tbt}}^{{#1}\lambda,\mathrm{rand}}}
\newcommand{\tbtdistx}{\mathbf{m}_{\mathrm{Tr,tbt}}^{\lambda,\mathrm{x}}}
\newcommand{\tbtpredistx}[1]{\mathbf{h}_{\mathrm{Tr,tbt}}^{{#1}\lambda,\mathrm{x}}}

\newcommand{\sqsupTr}{\sqsubseteq_{\mathrm{Tr,}\sqcup}^{\mathrm{det}}}
\newcommand{\supTr}{\sim_{\mathrm{Tr,}\sqcup}^{\mathrm{det}}}
\newcommand{\sqsupTrct}{\sqsubseteq_{\mathrm{Tr,}\sqcup}^{\mathrm{rand}}}
\newcommand{\supTrct}{\sim_{\mathrm{Tr,}\sqcup}^{\mathrm{rand}}}
\newcommand{\sqsupTrx}{\sqsubseteq_{\mathrm{Tr,}\sqcup}^{\mathrm{x}}}
\newcommand{\supTrx}{\sim_{\mathrm{Tr,}\sqcup}^{\mathrm{x}}}

\newcommand{\tracesupdist}{\mathbf{m}_{\mathrm{Tr,}\sqcup}^{\lambda,\mathrm{det}}}
\newcommand{\tracesuppredist}[1]{\mathbf{h}_{\mathrm{Tr,}\sqcup}^{{#1}\lambda,\mathrm{det}}}

\newcommand{\tracesupdistct}{\mathbf{m}_{\mathrm{Tr,}\sqcup}^{\lambda,\mathrm{rand}}}
\newcommand{\tracesuppredistct}[1]{\mathbf{h}_{\mathrm{Tr,}\sqcup}^{{#1}\lambda,\mathrm{rand}}}
\newcommand{\tracesupdistx}{\mathbf{m}_{\mathrm{Tr,}\sqcup}^{\lambda,\mathrm{x}}}
\newcommand{\tracesuppredistx}[1]{\mathbf{h}_{\mathrm{Tr,}\sqcup}^{{#1}\lambda,\mathrm{x}}}

\newcommand{\SC}{\mathbf{SC}}

\newcommand{\may}{\sim_{\mathrm{Te,may}}^{\mathrm{det}}}

\newcommand{\mayct}{\sim_{\mathrm{Te,may}}^{\mathrm{rand}}}
\newcommand{\sqmayx}{\sqsubseteq_{\mathrm{Te,may}}^{x}}
\newcommand{\mayx}{\sim_{\mathrm{Te,may}}^{x}}

\newcommand{\maydistx}{\mathbf{m}_{\mathrm{Te,may}}^{\omega,\mathrm{x}}}
\newcommand{\maypredistx}[1]{\mathbf{h}_{\mathrm{Te,may}}^{{#1}\omega,\mathrm{x}}}

\newcommand{\must}{\sim_{\mathrm{Te,must}}^{\mathrm{det}}}

\newcommand{\mustct}{\sim_{\mathrm{Te,must}}^{\mathrm{rand}}}
\newcommand{\sqmustx}{\sqsubseteq_{\mathrm{Te,must}}^{x}}
\newcommand{\mustx}{\sim_{\mathrm{Te,must}}^{x}}

\newcommand{\mustdistx}{\mathbf{m}_{\mathrm{Te,must}}^{\omega,\mathrm{x}}}
\newcommand{\mustpredistx}[1]{\mathbf{h}_{\mathrm{Te,must}}^{{#1}\omega,\mathrm{x}}}

\newcommand{\maymust}{\sim_{\mathrm{Te,mM}}^{\mathrm{det}}}

\newcommand{\maymustct}{\sim_{\mathrm{Te,mM}}^{\mathrm{rand}}}
\newcommand{\sqmaymustx}{\sqsubseteq_{\mathrm{Te,mM}}^{x}}
\newcommand{\maymustx}{\sim_{\mathrm{Te,mM}}^{x}}

\newcommand{\maymustdistx}{\mathbf{m}_{\mathrm{Te,mM}}^{\omega,\mathrm{x}}}
\newcommand{\maymustpredistx}[1]{\mathbf{h}_{\mathrm{Te,mM}}^{{#1}\omega,\mathrm{x}}}

\newcommand{\tbtTe}{\sim_{\mathrm{Te,tbt}}^{\mathrm{det}}}

\newcommand{\tbtTect}{\sim_{\mathrm{Te,tbt}}^{\mathrm{rand}}}
\newcommand{\sqtbtTex}{\sqsubseteq_{\mathrm{Te,tbt}}^{x}}
\newcommand{\tbtTex}{\sim_{\mathrm{Te,tbt}}^{x}}

\newcommand{\testtbtdist}{\mathbf{m}_{\mathrm{Te,tbt}}^{\lambda,\mathrm{det}}}
\newcommand{\testtbtpredist}[1]{\mathbf{h}_{\mathrm{Te,tbt}}^{{#1}\lambda,\mathrm{det}}}
\newcommand{\testtbtdistct}{\mathbf{m}_{\mathrm{Te,tbt}}^{\lambda,\mathrm{rand}}}
\newcommand{\testtbtpredistct}[1]{\mathbf{h}_{\mathrm{Te,tbt}}^{{#1}\lambda,\mathrm{rand}}}
\newcommand{\testtbtdistx}{\mathbf{m}_{\mathrm{Te,tbt}}^{\lambda,\mathrm{x}}}
\newcommand{\testtbtpredistx}[1]{\mathbf{h}_{\mathrm{Te,tbt}}^{{#1}\lambda,\mathrm{x}}}

\newcommand{\sqsupTe}{\sqsubseteq_{\mathrm{Te,}\sqcup}^{\mathrm{det}}}
\newcommand{\supTe}{\sim_{\mathrm{Te,}\sqcup}^{\mathrm{det}}}
\newcommand{\sqsupTect}{\sqsubseteq_{\mathrm{Te,}\sqcup}^{\mathrm{rand}}}
\newcommand{\supTect}{\sim_{\mathrm{Te,}\sqcup}^{\mathrm{rand}}}
\newcommand{\sqsupTex}{\sqsubseteq_{\mathrm{Te,}\sqcup}^{\mathrm{x}}}
\newcommand{\supTex}{\sim_{\mathrm{Te,}\sqcup}^{\mathrm{x}}}

\newcommand{\testsupdist}{\mathbf{m}_{\mathrm{Te,}\sqcup}^{\lambda,\mathrm{det}}}
\newcommand{\testsuppredist}[1]{\mathbf{h}_{\mathrm{Te,}\sqcup}^{{#1}\lambda,\mathrm{det}}}
\newcommand{\testsupdistct}{\mathbf{m}_{\mathrm{Te,}\sqcup}^{\lambda,\mathrm{rand}}}
\newcommand{\testsuppredistct}[1]{\mathbf{h}_{\mathrm{Te,}\sqcup}^{{#1}\lambda,\mathrm{rand}}}
\newcommand{\testsupdistx}{\mathbf{m}_{\mathrm{Te,}\sqcup}^{\lambda,\mathrm{x}}}
\newcommand{\testsuppredistx}[1]{\mathbf{h}_{\mathrm{Te,}\sqcup}^{{#1}\lambda,\mathrm{x}}}

\newcommand{\C}{\mathbf C}

\newcommand{\e}{\mathfrak{e}}

\newcommand{\OO}{\mathbf{O}}

\newcommand{\N}{\mathbb{N}} % Naturals
\newenvironment{apx-proof}[1] 
        {\noindent \textbf{Proof of #1.}} 
        {\qed}

\definecolor{lightblue}{RGB}{224,224,255}
\definecolor{lightred}{RGB}{255,224,224}
\definecolor{lightgreen}{RGB}{224,255,224}
\definecolor{lightyellow}{RGB}{255,255,224}
\definecolor{lightpurple}{RGB}{255,224,255}
\definecolor{darkerred}{RGB}{64,0,0}
\definecolor{darkred}{RGB}{128,0,0}
\definecolor{darkblue}{RGB}{0,0,128}
\definecolor{darkgreen}{RGB}{0,128,0}
\definecolor{darkpurple}{RGB}{128,0,128}
\definecolor{grey}{rgb}{0.745098,0.745098,0.745098}
\definecolor{lightgrey}{rgb}{0.9,0.9,0.9}
\definecolor{darkgrey}{rgb}{0.6,0.6,0.6}

\newcommand{\colorpar}[3]{\colorbox{#1}{\parbox{#2}{#3}}}

\newcommand{\marginremark}[3]{\marginpar{\colorpar{#2}{\linewidth}{\color{#1}#3}}}

\makeatletter
\def\THICKhrulefill{\leavevmode \leaders \hrule height 5pt\hfill \kern \z@}
\makeatother

\newcommand{\remarkML}[1]{\marginremark{darkblue}{lightblue}{\tiny{[ML]~ #1}}}
\newcommand{\remarkST}[1]{\marginremark{darkgreen}{lightgreen}{\tiny{[ST]~ #1}}}
\newcommand{\remarkVC}[1]{\marginremark{darkblue}{lightyellow}{\tiny{[VC]~ #1}}}

\ifdraft
\else
\renewcommand{\remarkML}[1]{}
\newcommand{\remarkST}[1]{}
\newcommand{\remarkVC}[1]{}

\fi

\newtheorem{theorem}{Theorem}
\newtheorem{proposition}{Proposition}

\theoremstyle{definition}
\newtheorem{definition}{Definition}
\newtheorem{example}{\emph{Example}}
\let\doendexample\endexample
\renewcommand\endexample{~\hfill$\LHD$\doendexample} 

\theoremstyle{remark}
\newtheorem{remark}{Remark}

\title{Trace and Testing Metrics on Nondeterministic Probabilistic Processes}
\author{Valentina Castiglioni
\institute{INRIA Saclay - Ile de France, France}
\email{valentina.castiglioni@inria.fr}
}

\def\titlerunning{Trace and Testing Metrics on PTSs}
\def\authorrunning{V. Castiglioni}

\begin{document}

\maketitle

\begin{abstract}
The combination of \emph{nondeterminism and probability} in concurrent systems lead to the development of several interpretations of process behavior.
If we restrict our attention to linear properties only, we can identify three main approaches to \emph{trace} and \emph{testing} semantics: the \emph{trace distributions}, the \emph{trace-by-trace} and the \emph{extremal probabilities} approaches.
In this paper, we propose novel notions of \emph{behavioral metrics} that are based on the three classic approaches above, and that can be used to measure the disparities in the linear behavior of processes wrt.\ trace and testing semantics.
We study the properties of these metrics, like \emph{non-expansiveness}, and we compare their expressive powers.
\end{abstract}

%===================================================
% sec 1 - intro
%====================================================

\section{Introduction}

A major task in the development of complex systems is to verify that an \emph{implementation} of a system meets its \emph{specification}.
Typically, in the realm of process calculi, implementation and specification are \emph{processes} formalized with the same language, and the verification task consists in \emph{comparing their behavior}, which can be done at different levels of abstraction, depending on which aspects of the behavior can be ignored or must be captured.
If one focuses on linear properties only, processes are usually compared on the basis of the \emph{traces} they can execute, or accordingly to their capacity to pass the same \emph{tests}.
This was the main idea behind the study of \emph{trace equivalence}~\cite{H85} and \emph{testing equivalence}~\cite{dNH84}. 

If we consider also probabilistic aspects of system behavior, reasoning in terms of qualitative equivalences is only partially satisfactory. 
Any tiny variation of the probabilistic behavior of a system, which may be also due to a measurement error, will break the equality between processes without any further information on the \emph{distance} of their behaviors. 
Actually, many implementations can only \emph{approximate} the specification; thus, the verification task requires appropriate instruments to measure the \emph{quality} of the approximation. 
For this reason, we propose to use \emph{hemimetrics} measuring the disparities in process behavior wrt.\ linear semantics also to \emph{quantify} process verification.
Informally, we may see a specification not as the precise desired behavior of the system, but as set of \emph{minimum requirements} on system behavior, such as the \emph{lower bounds} on the probabilities to execute given traces or pass given tests.
Then, given a \emph{hemimetric} $\mathbf{h}$ expressing trace (resp.\ testing) semantics, we can set a certain \emph{tolerance} $\varepsilon$, related to the application context, and transform the verification problem into a \emph{verification up-to-}$\varepsilon$, or \emph{$\varepsilon$-robustness} problem: we say that an implementation $I$ is $\varepsilon$-\emph{trace-robust} (resp.\ $\varepsilon$-\emph{testing-robust}) wrt.\ a specification $S$ if whenever $S$ can perform a trace (resp.\ pass a test) with a given probability $p$, then $I$ can do the same with probability \emph{at least} $p-\varepsilon$, namely if $\mathbf{h}(S,I) \le \varepsilon$.
Dually, we may see $S$ as giving an \emph{upper bound} to undesired system behavior, and demand that whenever $S$ can perform a trace (resp.\ pass a test) with a given probability $p$, then $I$ can do the same with probability \emph{at most} $p+\varepsilon$, namely if $\mathbf{h}(I,S) \le \varepsilon$.

In this paper, we consider \emph{nondeterministic probabilistic labeled transition systems} (\emph{PTS}) \cite{S95}, a very general model in which \emph{nondeterminism and probability} coexist, and we discuss the definition of \emph{hemimetrics} and \emph{pseudometrics} suitable to measure the differences in process behavior wrt.\ trace and testing semantics.
We will see that the interplay of probability and nondeterminism lead to some difficulties in defining notions of behavioral distance, as already experienced in the case of equivalences~\cite{BdNL14}.
For instance, in trace semantics, it is questionable whether the choice of the trace should precede or follow the choice by the \emph{scheduler}. 

Several approaches to probabilistic trace equivalence are discussed in \cite{BdNL14}:
\begin{inparaenum}[(i)] 
\item \label{tdapp}
The \emph{trace distribution}~\cite{S95tr} approach, comparing \emph{entire resolutions} created by schedulers by checking if they assign the same probability to the same traces;
\item \label{tbtapp}
The \emph{trace-by-trace}~\cite{BdNL12} approach, in which firstly we take a trace and then we check if there are resolutions for processes assigning the same probability to it;
\item \label{epapp}
The \emph{extremal probabilities}~\cite{BdNL13} approach, considering for each trace only the infima and suprema of the probabilities assigned to it over all resolutions for the processes.
\end{inparaenum}
We will argue that considering only \emph{supremal probabilities} instead of both extremal probabilities is more tailored to reason on the verification problem.
Then, we propose three \emph{trace hemimetrics} and \emph{pseudometrics} as quantitative variants of trace distribution, trace-by-trace and supremal probabilities trace preorders and equivalences. 
All these distances are parametric wrt.\ the type of scheduler.
We consider \emph{deterministic} and \emph{randomized} schedulers, however an extension to other types of schedulers seems feasible.
Our results can be summarized as follows:
\begin{inparaenum}
\item We prove that, under each hemimetric/pseudometric, the pairs of processes at distance zero are precisely those related by the corresponding preorder/equivalence.
\item We prove that the hemimetrics/pseudometrics for trace-by-trace and supremal probabilities semantics are suitable for compositional reasoning, by showing their \emph{non-expansiveness}~\cite{DGJP04} wrt.\ parallel composition. 
\item We study the dif{f}erences in the expressive powers of these distances, thus composing them in a simple spectrum.
In particular, we show that the supremal probabilities semantics defined either on deterministic or randomized schedulers has the same expressive power of the trace-by-trace semantics on randomized schedulers.
This is a very interesting result in the perspective of an application to quantitative process verification: the comparison of the suprema execution probabilities of linear properties has the same expressive power of a pairwise comparison of the probabilities in all possible randomized resolutions of nondeterminism.
\end{inparaenum}

Then, we consider three approaches to testing semantics:
\begin{inparaenum}[(i)]
\item 
\label{maymustapp}
the \emph{may/must}~\cite{YL92},
\item 
\label{tbttestingapp}
the \emph{trace-by-trace}~\cite{BdNL14}, 
\item 
\label{suptestingapp}
the \emph{supremal probabilities} approach.
\end{inparaenum}
Briefly, in (\ref{maymustapp}) the extremal probabilities of passing a test are considered whereas (\ref{tbttestingapp})--(\ref{suptestingapp}) base on a \emph{traced} view of testing, in that we compare the probabilities of passing the test via the execution of a given trace.
Actually, (\ref{tbttestingapp})--(\ref{suptestingapp})  can be considered as the adaptation to testing semantics of the trace-by-trace and suprema probability approaches to trace semantics. 
For each of these approaches, we present a hemimetric and a pseudometric as the quantitative variant of the related preorder and equivalence.
To the best of our knowledge, ours is the first attempt in this direction. 
In detail: 
\begin{inparaenum}
\item We prove that, under each hemimetric/pseudometric, the pairs of processes at distance zero are precisely those equated by the related testing preorder/equivalence.
\item We prove that all hemimetrics and pseudometrics are non-expansive.
\item We compose these testing distances in a simple spectrum and we also compare them with trace distances.
\end{inparaenum}

%=======================================================
% Back
%========================================================

\section{Background}
\label{sec:back}

PTSs \cite{S95} are a very general model combining LTSs \cite{K76} and discrete time Markov chains \cite{HJ94}, to model reactive behavior, nondeterminism and probability.
In a PTS, the state space is given by a set $\proc$ of $\emph{processes}$, ranged over by  $s,t,\dots$ and transition steps take processes to \emph{probability distributions} over processes.
Probability distributions over $\proc$ are mappings $\pi \colon \proc \to [0,1]$ with $\sum_{s \in \proc} \pi(s) = 1$.
By $\ProbDist{\proc}$ we denote the set of all distributions over $\proc$, ranged over by $\pi,\pi',\dots$
For $\pi \in \ProbDist{\proc}$, the \emph{support} of $\pi$ is the set $\support(\pi) = \{ s \in \proc \mid \pi(s) >0\}$.
We consider only distributions with \emph{finite} support.
For $s \in \proc$, we let $\delta_s$ denote the~\emph{Dirac distribution on $s$} defined by $\delta_s(s)= 1$ and $\delta_s(t)=0$ for $t\neq s$.

\begin{definition}[PTS, \cite{S95}]
A \emph{nondeterministic probabilistic labeled transition system (PTS)} is a triple $(\proc,\Act,\trans[])$ where: 
\begin{inparaenum}[(i)]
\item $\proc$ is a countable set of processes, 
\item $\Act$ is a countable set of \emph{actions}, and 
\item $\trans[] \subseteq {\proc \times \Act \times \ProbDist{\proc}}$ is a \emph{transition relation}. 
\end{inparaenum}
\end{definition}

We write $s\trans[a]\pi$ for $(s,a,\pi) \in\trans[]$, $s \trans[a] $ if there is a distribution $\pi$ with $s \trans[a] \pi$, and $s \ntrans[a]$ otherwise. 
A PTS is \emph{fully nondeterministic} if every transition has the form $s\trans[a]\delta_t$ for some $t \in \proc$.
A PTS is \emph{fully probabilistic} if at most one transition is enabled for each process.
$s \in \proc$ is \emph{image-finite} \cite{HPSWZ11} if for each $a \in \Act$ the number of $a$-labeled transitions enabled for $s$ is finite.
We consider only image-finite processes.

\begin{definition}
[Parallel composition]
Let $P_1 = (S_1, \Act, \trans[]_1)$ and $P_2 = (S_2, \Act, \trans[]_2)$ be two PTSs.
The (CSP-like \cite{H85})
\emph{synchronous parallel composition of $P_1$ and $P_2$} is the PTS $P_1 \parallel P_2 = (S_1 \times S_2, \Act, \trans[])$, where $\trans[] \subseteq (S_1 \times S_2) \times \Act \times \ProbDist{S_1 \times S_2}$ is such that $(s_1,s_2) \trans[a] \pi$ if and only if $s_1 \trans[a]_1 \pi_1$, $s_2 \trans[a]_2 \pi_2$ and $\pi(s_1',s_2') = \pi_1(s'_1) \cdot \pi_2(s'_2)$ for all $(s'_1,s'_2) \in S_1 \times S_2$.
\end{definition}

%==================================================

We proceed to recall some notions, mostly from \cite{BdNL13,BdNL14,BdNL14bis}, necessary to reason on trace and testing semantics.
A \emph{computation} is a weighted sequence of process-to-process transitions.

\begin{definition}
[Computation]
\label{def:computation}
A  \emph{computation from $s_0$ to $s_n$} has the form\\
$\begin{array}{c}
\hspace{3.5 cm }c := s_0 \ctrans[a_1] s_1 \ctrans[a_2] s_2 \dots s_{n-1} \ctrans[a_n] s_n
\end{array}
$\\
where, for all $i = 1,\dots,n$, there is a transition $s_{i-1} \trans[a_i] \pi_i$ with $s_i \in \support(\pi_i)$.
\end{definition}
Note that $\pi_i(s_i)$ is the \emph{execution probability} of step $s_{i-1} \ctrans[a_i] s_i$ conditioned on the selection of the transition $s_{i-1} \trans[a_i] \pi_i$ at $s_{i-1}$.
We denote by $\pr(c) = \prod_{i = 1}^{n} \pi_i(s_i)$ the product of the execution probabilities of the steps in $c$.
A computation $c$ from $s$ is \emph{maximal} if it is not a proper prefix of any other computation from $s$.
We denote by $\C(s)$ (resp.\ $\C_{\max}(s)$) the set of computations (resp.\ maximal computations) from $s$.
For any $\C \subseteq \C(s)$, we define $\pr(\C) = \sum_{c \in \C} \pr(c)$ whenever none of the computations in $\C$ is a proper prefix of any of the others.

We denote by $\Act^{\star}$ the set of \emph{finite traces} in $\Act$ and write $\e$ for the empty trace.
We say that a computation is \emph{compatible} with the trace $\alpha \in \Act^{\star}$ if{f} the sequence of actions labeling the computation steps is equal to $\alpha$.
We denote by $\C(s,\alpha) \subseteq \C(s)$ the set of computations from $s$ that are compatible with $\alpha$, and by $\C_{\max}(s,\alpha)$ the set $\C_{\max}(s,\alpha) = \C_{\max}(s) \cap \C(s,\alpha)$.

To express linear semantics we need to evaluate and compare the probability of particular sequences of \emph{events} to occur.
As in PTSs this probability highly depends also on nondeterminism, \emph{schedulers} \cite{S95tr,WJ06,GA12} (or \emph{adversaries}) resolving it become fundamental.
They can be classified into two main classes: \emph{deterministic} and \emph{randomized schedulers} \cite{S95tr}.
For each process, a deterministic scheduler selects exactly one transition among the possible ones, or none of them, thus treating all internal nondeterministic choices as distinct.
Randomized schedulers allow for a convex combination of the equally labeled transitions.
The resolution given by a deterministic scheduler is a fully probabilistic process, whereas from randomized schedulers we get a fully probabilistic process with \emph{combined transitions} \cite{SL95}.

\begin{definition}
[Resolutions]
\label{def:det_res}
Let $P = (\proc, \Act,\trans[])$ be a PTS and $s \in \proc$.
We say that a PTS $\Z = (Z,\Act,\trans[]_{\Z})$ is a \emph{deterministic resolution} for $s$ if{f} there exists a function $\corr{\Z} \colon Z \to \proc$ such that $s = \corr{\Z}(z_s)$ for some $z_s \in Z$ and moreover:
\begin{compactitem}
\item[(i)] \label{item:res_same_distributions}
If $z \trans[a]_{\Z} \pi$, then $\corr{\Z}(z) \trans[a] \pi'$ with $\pi(z') = \pi'(\corr{\Z}(z'))$ for all $z' \in Z$.
\item[(ii)] If $z \trans[a_1]_{\Z} \pi_1$ and $z \trans[a_2]_{\Z} \pi_2$ then $a_1 = a_2$ and $\pi_1 = \pi_2$. 
\end{compactitem}
Conversely, we say that $\Z$ is a \emph{randomized resolution} for $s$ if item (i) is replaced by
\begin{compactitem}
\item[(i)'] If $z \trans[a]_{\Z} \pi$, then there are $n \in \N$, $\{p_i \in (0,1] \mid \sum_{i = 1}^n p_i = 1\}$ and $\{\corr{\Z}(z) \trans[a] \pi_i \mid 1 \le i \le n\}$ s.t.\ $\pi(z') = \sum_{i = 1}^n p_i \cdot \pi_i(\corr{\Z}(z'))$ for all $z' \in Z$.
\end{compactitem}
Then, $\Z$ is \emph{maximal} if{f} it cannot be further extended in accordance with the graph structure of $P$ and the constraints above.
For $\mathrm{x} \in \{\mathrm{det},\mathrm{rand}\}$, we denote by $\resx(s)$ the set of resolutions for $s$ and by $\resx_{\max}(s)$ the subset of maximal resolutions for $s$.
\end{definition}

We conclude this section by recalling the mathematical notions of hemimetric and pseudometric.
A $1$-bounded \emph{pseudometric} on $\proc$ is a function $d \colon \proc \times \proc \to [0,1]$ s.t.:  
\begin{inparaenum}[(i)]
\item \label{zero}
$d(s,s) =0$,
\item \label{symm}
 $d(s,t) = d(t,s)$, 
\item \label{tri}
 $d(s,t) \le d(s,u) + d(u,t)$, 
\end{inparaenum}
for $s,t,u \in \proc$.
Then, $d$ is a \emph{hemimetric} if it satisfies (\ref{zero}) and (\ref{tri}).
The \emph{kernel} of a (hemi,pseudo)metric $d$ on $\proc$ the set of pairs of elements in $\proc$ which are at distance $0$, namely $ker(d) = \{(s,t) \in \proc \times \proc \mid d(s,t) = 0\}$.

\emph{Non-expansiveness} \cite{DGJP04} of a (hemi,pseudo)metric is the quantitative analogue to the (pre)congruence property.
Here we propose also a stronger notion, called \emph{strict non-expansiveness} that gives tighter bounds on the distance of processes composed in parallel.

\begin{definition}
[(Strict) non-expansiveness]
\label{def:non_expansiveness}
Let $d$ be a (hemi,pseudo)metric on $\proc$.
Following \cite{DGJP04}, we say that $d$ is \emph{non-expansive} wrt.\ the parallel composition operator if and only if for all $s_1,s_2,t_1,t_2 \in \proc$ we have
$d(s_1 \parallel s_2, t_1 \parallel t_2) \le d(s_1,t_1) + d(s_2,t_2)$.
Moreover, we say that $d$ is \emph{strictly non-expansive} if $d(s_1 \parallel s_2, t_1 \parallel t_2) \le d(s_1,t_1) + d(s_2,t_2) - d(s_1,t_1) \cdot d(s_2,t_2)$.
\end{definition}

Finally, we remark that, as elsewhere in the literature, throughout the paper we may use the term \emph{metric} in place of pseudometric.

%========================================================
% Traces
%=========================================================

\section{Metrics for traces}
\label{sec:trace_metric}

In this Section, we define the metrics measuring the disparities in process behavior wrt.\ trace semantics.
We consider three approaches to the combination of nondeterminism and probability: the \emph{trace distribution}, the \emph{trace-by-trace} and the \emph{supremal probabilities} approach.

In defining the behavioral distances, we assume a \emph{discount factor $\lambda \in (0,1]$}, which allows us to specify how much the behavioral distance of future transitions is taken into account~\cite{AHM03,DGJP04}. 
The discount factor $\lambda=1$ expresses no discount, so that the dif{f}erences in the behavior between $s,t \in \proc$ are considered irrespective of after how many steps they can be observed.

%==============================================

\subsection{The trace distribution approach}

In \cite{S95tr} the observable events characterizing the trace semantics are \emph{trace distributions}, ie.\ probability distributions over traces.
Processes $s,t \in\proc$ are \emph{trace distribution equivalent} if, \emph{for any resolution} for $s$ there is a resolution for $t$ exhibiting \emph{the same trace distribution}, ie.\ the execution probability of each trace in the two resolutions is \emph{exactly the same}, and vice versa.

\begin{definition}
[Trace distribution equivalence \cite{S95tr,BdNL12}]
\label{def:trdist}
Let $(\proc,\Act,\trans)$ be a PTS and $\mathrm{x} \in \{\mathrm{det},\mathrm{rand}\}$.
Processes $s,t \in \proc$ are in the \emph{trace distribution preorder}, written $s \sqdistx t$, if: 
\begin{center}
$\text{for each } \Z_s \in \resx(s) \text{ there is } \Z_t \in \resx(t) \text{ s.t.\ for each } \alpha \in \Act^{\star} \colon \pr(\C(z_s,\alpha)) = \pr(\C(z_t,\alpha))$. 
\end{center}
Then, $s,t$ are \emph{trace distribution equivalent}, notation $s \trdistx t$, if{f} $s \sqdistx t$ and $t \sqdistx s$.
\end{definition}

The quantitative analogue to trace distribution equivalence is based on the evaluation of the \emph{dif{f}erences in the trace distributions} of processes: the distance between processes $s,t$ is set to $\varepsilon \ge 0$ if, \emph{for any resolution} for $s$ there is a resolution for $t$ exhibiting \emph{a trace distribution differing at most by $\varepsilon$}, meaning that the execution probability of each trace in the two resolutions \emph{dif{f}ers by at most $\varepsilon$}, and vice versa.

\begin{definition}
[Trace distribution metric]
\label{def:trdistdist}
Let $(\proc,\Act,\trans)$ be a PTS, $\lambda \in (0,1]$ and $\mathrm{x} \in \{\mathrm{det},\mathrm{rand}\}$. 
The \emph{trace distribution hemimetric} and the \emph{trace distribution metric} are the functions $\trdistpredistx, \trdistdistx\colon \proc \times \proc \to [0,1]$ defined for all $s,t \in \proc$ by
\begin{compactitem}
\item $\trdistpredistx(s,t) = \sup_{\Z_s \in \resx(s)}\, \inf_{\Z_t \in \resx(t)}\, \sup_{\alpha \in \Act^{\star}} \lambda^{|\alpha|-1}|\pr(\C(z_s,\alpha)) - \pr(\C(z_t,\alpha))|$
\item $\trdistdistx(s,t) = \max\{\trdistpredistx(s,t),\trdistpredistx(t,s)\}$.
\end{compactitem}
\end{definition}

We observe that the expression $\sup_{\alpha \in \Act^{\star}} \lambda^{|\alpha|-1}|\pr(\C(z_s,\alpha)) - \pr(\C(z_t,\alpha))|$ 
used in Definition~\ref{def:trdistdist} corresponds to the (weighted) total variation distance between the trace distributions given by the two resolutions $\Z_s$ and $\Z_t$.
An equivalent formulation is given in \cite{SDC07,CT17} via the Kantorovich lifting of the discrete metric over traces.

We now state that trace distribution hemimetrics and metrics are well-defined and that their kernels are the trace distribution preorders and equivalences, respectively.

\begin{theorem}
\label{thm:trdistdist_is_metric}
Let $(\proc,\Act,\trans)$ be a PTS, $\lambda \in (0,1]$ and $\mathrm{x} \in \{\mathrm{det},\mathrm{rand}\}$. 
Then:
\begin{enumerate}
\item The function $\trdistpredistx$ is a $1$-bounded hemimetric on $\proc$, with $\sqdistx$ as kernel.
\item The function $\trdistdistx$ is a $1$-bounded pseudometric on $\proc$, with $\trdistx$ as kernel.
\end{enumerate}
\end{theorem}

\begin{figure}[t!]
\centering
\begin{tikzpicture}
\node at (2.4,2.1){$\boldsymbol{s_p}$};
\draw[->](2.4,1.9)--(0.6,1.4);
\node at (1,1.8){\scalebox{0.8}{$a$}};
\draw[->](2.4,1.9)--(2.4,1.4);
\node at (2.6,1.6){\scalebox{0.8}{$a$}};
\draw[->](2.4,1.9)--(4.2,1.4);
\node at (3.8,1.8){\scalebox{0.8}{$a$}};
\draw[->,dotted,thick](0.6,1.4)--(0,0.9);
\node at (0,1.2){\scalebox{0.8}{$p$}};
\draw[->,dotted,thick](0.6,1.4)--(1.2,0.9);
\node at (1.3,1.2){\scalebox{0.8}{$1$-$p$}};
\draw[->,dotted,thick](2.4,1.4)--(2.4,0.9);
\node at (2.6,1.2){\scalebox{0.8}{$1$}};
\draw[->,dotted,thick](4.2,1.4)--(3.6,0.9);
\node at (3.4,1.2){\scalebox{0.8}{$1$-$p$}};
\draw[->,dotted,thick](4.2,1.4)--(4.8,0.9);
\node at (4.8,1.2){\scalebox{0.8}{$p$}};
\node at (0,0.7){$\boldsymbol{s_1}$};
\node at (1.2,0.7){$\boldsymbol{\nil}$};
\node at (2.4,0.7){$\boldsymbol{s_2}$};
\node at (3.6,0.7){$\boldsymbol{\nil}$};
\node at (4.8,0.7){$\boldsymbol{s_3}$};
\draw[->](0,0.5)--(0,0);
\node at (0.2,0.3){\scalebox{0.8}{$b$}};
\draw[->](2.4,0.5)--(1.8,0);
\node at (1.9,0.3){\scalebox{0.8}{$b$}};
\draw[->](2.4,0.5)--(3,0);
\node at (2.9,0.3){\scalebox{0.8}{$c$}};
\draw[->](4.8,0.5)--(4.8,0);
\node at (5,0.3){\scalebox{0.8}{$c$}};
\node at (7.2,2.1){$\boldsymbol{t}$};
\draw[->](7.2,1.9)--(6,1.4);
\node at (6.2,1.8){\scalebox{0.8}{$a$}};
\draw[->](7.2,1.9)--(7.2,1.4);
\node at (7.4,1.6){\scalebox{0.8}{$a$}};
\draw[->](7.2,1.9)--(8.4,1.4);
\node at (8.2,1.8){\scalebox{0.8}{$a$}};
\draw[->,dotted,thick](6,1.4)--(6,0.9);
\node at (5.8,1.2){\scalebox{0.8}{$1$}};
\draw[->,dotted,thick](7.2,1.4)--(6.6,0.9);
\node at (6.6,1.2){\scalebox{0.8}{$0.5$}};
\draw[->,dotted,thick](7.2,1.4)--(7.8,0.9);
\node at (7.8,1.2){\scalebox{0.8}{$0.5$}};
\draw[->,dotted,thick](8.4,1.4)--(8.4,0.9);
\node at (8.6,1.2){\scalebox{0.8}{$1$}};
\node at (6,0.7){$\boldsymbol{t_1}$};
\node at (6.6,0.7){$\boldsymbol{t_2}$};
\node at (7.8,0.7){$\boldsymbol{t_3}$};
\node at (8.4,0.7){$\boldsymbol{t_4}$};
\draw[->](6,0.5)--(6,0);
\node at (6.2,0.3){\scalebox{0.8}{$b$}};
\draw[->](6.6,0.5)--(6.6,0);
\node at (6.8,0.3){\scalebox{0.8}{$b$}};
\draw[->](7.8,0.5)--(7.8,0);
\node at (8,0.3){\scalebox{0.8}{$c$}};
\draw[->](8.4,0.5)--(8.4,0);
\node at (8.6,0.3){\scalebox{0.8}{$c$}};
\node at (9.6,2.1){$\boldsymbol{z_{t}^1}$};
\draw[->](9.6,1.9)--(9.6,1.4);
\node at (9.8,1.7){\scalebox{0.8}{$a$}};
\draw[->,dotted,thick](9.6,1.4)--(9.6,1);
\node at (9.8,1.2){\scalebox{0.8}{$1$}};
\node at (9.6,0.7){$\boldsymbol{z_{t_1}^1}$};
\node at (11.4,2.1){$\boldsymbol{z_t^2}$};
\draw[->](11.4,1.9)--(11.4,1.4);
\node at (11.6,1.7){\scalebox{0.8}{$a$}};
\draw[->,dotted,thick](11.4,1.4)--(10.8,1);
\node at (10.8,1.3){\scalebox{0.8}{$0.5$}};
\draw[->,dotted,thick](11.4,1.4)--(12,1);
\node at (12,1.3){\scalebox{0.8}{$0.5$}};
\node at (10.8,0.7){$\boldsymbol{z_{t_2}^2}$};
\node at (12,0.7){$\boldsymbol{z_{t_3}^2}$};
\draw[->](10.8,0.5)--(10.8,0);
\node at (11,0.3){\scalebox{0.8}{$b$}};
\node at (13,2.1){$\boldsymbol{z_{t}^3}$};
\draw[->](13,1.9)--(13,1.4);
\node at (13.2,1.7){\scalebox{0.8}{$a$}};
\draw[->,dotted,thick](13,1.4)--(13,1);
\node at (13.2,1.2){\scalebox{0.8}{$1$}};
\node at (13,0.7){$\boldsymbol{z_{t_1}^3}$};
\draw[->](13,0.5)--(13,0);
\node at (13.2,0.3){\scalebox{0.8}{$b$}};
\end{tikzpicture}
\caption{\label{fig:one_for_all} We will evaluate the trace distances between $s_p$ and $t$ wrt.\ the dif{f}erent approaches, schedulers and  parameter $p \in [0,1]$.
In all upcoming examples we will investigate only the traces that are significant for the evaluation of the considered distance.}
\end{figure}
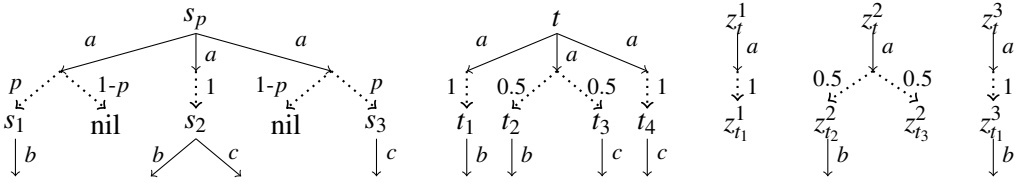

\begin{example}
\label{ex:trdist}
Consider processes $s_p$ and $t$ in Figure~\ref{fig:one_for_all}, with $p \in [0,1]$.
First we evaluate $\trdistpredist(t,s_p)$.
We expand only the case for the resolution $\Z_t$ for $t$ obtained from its central $a$-branch.
It assigns probability $0.5$ to both $ab$ and $ac$.
Under deterministic schedulers, any resolution $\Z_{s_p}$ for $s_p$ can assign positive probability to only one of these traces.
Assume this trace is $ab$, the case $ac$ is analogous.
We have either $\pr(\C(z_{s_p},ab)) = p$ or $\pr(\C(z_{s_p},ab)) = 1$.
Then, $|\pr(\C(z_{t},ab)) - \pr(\C(z_{s_p},ab))| \in\{0.5,|0.5-p|\}$ and $|\pr(\C(z_{t},ac)) - \pr(\C(z_{s_p},ac))| = |0.5-0|=0.5$.
Therefore, $\trdistpredist(t,s_p) = \lambda \cdot 0.5$, for all $p \in [0,1]$.

Now we show that $\trdistpredist(s_p,t) = \lambda \cdot \min\{p, |0.5 - p|,1-p\}$.
For each resolution $\Z_{s_p}$ for $s_p$ we need the resolution for $t$ whose trace distribution is closer to that of $\Z_{s_p}$.
We expand only the case of $\Z_{s_p}$ corresponding to the leftmost $a$-branch of $s_p$ and giving probability $1$ to trace $a$ and $p$ to trace $ab$.
We distinguish three subcases, related to the value of $p$:
\begin{inparaenum}[(i)]
\item $p \in [0,0.25]$:
The resolution for $t$ minimizing the distance from $\Z_{s_p}$ is $\Z_t^1$ that selects no action for $z^1_{t_1}$. 
The distance between $\Z_{s_p}$ and $\Z_t^1$ is $\lambda^{|ab|-1}|\pr(\C(z_{s_p},ab)) - \pr(\C(z_t^1,ab))| = \lambda\cdot p$.
Notice that in this case $p \le |0.5-p|,1-p$.
\item $p \in (0.25,0.75]$: 
The resolution for $t$ that minimizes the distance from $\Z_{s_p}$ is $\Z_t^2$ that performs an $a$-move and evolves to $0.5 \delta_{z^2_{t_2}} + 0.5 \delta_{z^2_{t_3}}$, where $z^2_{t_3}$ that executes no action.
The distance between $\Z_{s_p}$ and $\Z^2_t$ is $\lambda^{|ab|-1}|\pr(\C(z_{s_p},ab)) - \pr(\C(z^2_t,ab))| = \lambda\cdot |0.5 - p|$.
Notice that in this case we have $|0.5-p| \le p,1-p$.
\item $p \in (0.75,1]$:
The resolution for $t$ that minimizes the distance from $\Z_{s_p}$ is $\Z_t^3$ that corresponds to the leftmost branch of $t$. 
The distance between $\Z_s$ and $\Z^3_t$ is $\lambda^{|ab|-1}|\pr(\C(z_{s_p},ab)) - \pr(\C(z^3_t,ab))| = \lambda\cdot (1-p)$.
Notice that in this case $1-p \le |0.5-p|,p$.
\end{inparaenum}  

In the case of randomized schedulers, one can prove that, since both $s_p,t$ can perform traces $ab$ and $ac$ with probability $1$, for any $p \in [0,1]$ we get $\trdistpredistct(s_p,t) = \trdistpredistct(t,s_p) =0$.
\end{example}

%===========================================

\subsection{The trace-by-trace approach}
\label{sec:trace_tbt}

Trace distribution equivalences come with some desirable properties, as the full backward compatibility with the fully nondeterministic and fully probabilistic cases (cf.\ \cite[Thm.~3.4]{BdNL14}).
However, they are not congruences wrt.\ parallel composition \cite{S95tr}, and thus the related metrics cannot be non-expansive.
Moreover, due to the crucial r\^ole of the schedulers in the discrimination process, trace distribution distances are sometimes too demanding.
Take, for example, processes $s,t$ in Figure~\ref{fig:tbt-vs-td}, with $\varepsilon_1,\varepsilon_2 \in [0,0.5]$.
We have $\trdistpredist(s,t) = \lambda \cdot 0.5$ and $\trdistpredist(t,s) = \lambda \cdot \max_{i \in \{1,2\}} \max\{0.5- \varepsilon_i, \varepsilon_i\}$, thus giving $\trdistdist(s,t) = \lambda \cdot 0.5$ for all $\varepsilon_1,\varepsilon_2 \in [0,0.5]$. 
However, $s$ and $t$ can perform the same traces with probabilities that differ at most by $\max(\varepsilon_1,\varepsilon_2)$, which suggests that their trace distance should be $\lambda \cdot \max(\varepsilon_1,\varepsilon_2)$.
Specially, for $\varepsilon_1,\varepsilon_2=0$, $s,t$ can perform the same traces with exactly the same probability.
Despite this, $s,t$ are still distinguished by trace distribution equivalences.
These situations arise since the focus of trace distribution approach is more on resolutions than on traces.

To move the focus on traces, the \emph{trace-by-trace} approach was proposed \cite{BdNL12}.
The idea is to choose first the event that we want to observe, namely \emph{a single trace}, and only as a second step we let the scheduler perform its selection: processes $s,t$ are equivalent wrt.\ the trace-by-trace approach if \emph{for each trace} $\alpha$, for each resolution for $s$ there is a resolution for $t$ that assigns to $\alpha$ the \emph{same probability}, and vice versa.

\begin{figure}[t!]
\centering
\begin{tikzpicture}
\node at (3,4.5){$\boldsymbol{s}$};
\draw[->](3,4.3)--(1.4,3.8);
\node at (2.2,4.2){\scalebox{0.8}{$a$}};
\draw[->](3,4.3)--(4.6,3.8);
\node at (3.8,4.2){\scalebox{0.8}{$a$}};
\draw[dotted,thick,->](1.4,3.8)--(0.6,3.3);
\node at (0.7,3.65){\scalebox{0.8}{$0.5$}};
\draw[dotted,thick,->](1.4,3.8)--(2.2,3.3);
\node at (2.1,3.65){\scalebox{0.8}{$0.5$}};
\draw[dotted,thick,->](4.6,3.8)--(3.8,3.3);
\node at (3.9,3.65){\scalebox{0.8}{$0.5$}};
\draw[dotted,thick,->](4.6,3.8)--(5.4,3.3);
\node at (5.3,3.65){\scalebox{0.8}{$0.5$}};
\node at (0.6,3.1){$\boldsymbol{s_1}$};
\node at (2.2,3.1){$\boldsymbol{s_2}$};
\node at (3.8,3.1){$\boldsymbol{s_3}$};
\node at (5.4,3.1){$\boldsymbol{s_4}$};
\draw[->](0.6,2.9)--(0.6,2.4);
\node at (0.4,2.7){\scalebox{0.8}{$b$}};
\draw[->](2.2,2.9)--(2.2,2.4);
\node at (2.4,2.7){\scalebox{0.8}{$c$}};
\draw[->](3.8,2.9)--(3.8,2.4);
\node at (3.6,2.7){\scalebox{0.8}{$d$}};
\draw[->](5.4,2.9)--(5.4,2.4);
\node at (5.6,2.7){\scalebox{0.8}{$e$}};
\node at (10,4.5){$\boldsymbol{t}$};
\draw[->](10,4.3)--(8.4,3.8);
\node at (9.2,4.2){\scalebox{0.8}{$a$}};
\draw[->](10,4.3)--(11.6,3.8);
\node at (10.8,4.2){\scalebox{0.8}{$a$}};
\draw[dotted,thick,->](8.4,3.8)--(7.6,3.3);
\node at (7.5,3.65){\scalebox{0.8}{$0.5$-$\varepsilon_1$}};
\draw[dotted,thick,->](8.4,3.8)--(9.2,3.3);
\node at (9.25,3.65){\scalebox{0.8}{$0.5$+$\varepsilon_1$}};
\draw[dotted,thick,->](11.6,3.8)--(10.8,3.3);
\node at (10.75,3.65){\scalebox{0.8}{$0.5$-$\varepsilon_2$}};
\draw[dotted,thick,->](11.6,3.8)--(12.4,3.3);
\node at (12.5,3.65){\scalebox{0.8}{$0.5$+$\varepsilon_2$}};
\node at (7.6,3.1){$\boldsymbol{t_1}$};
\node at (9.2,3.1){$\boldsymbol{t_2}$};
\node at (10.8,3.1){$\boldsymbol{t_3}$};
\node at (12.4,3.1){$\boldsymbol{t_4}$};
\draw[->](7.6,2.9)--(7.6,2.4);
\node at (7.4,2.7){\scalebox{0.8}{$b$}};
\draw[->](9.2,2.9)--(9.2,2.4);
\node at (9.4,2.7){\scalebox{0.8}{$d$}};
\draw[->](10.8,2.9)--(10.8,2.4);
\node at (10.6,2.7){\scalebox{0.8}{$c$}};
\draw[->](12.4,2.9)--(12.4,2.4);
\node at (12.6,2.7){\scalebox{0.8}{$e$}};
\end{tikzpicture}
\caption{\label{fig:tbt-vs-td} For $\varepsilon_1,\varepsilon_2 \in [0,0.5]$, we have $\tbtdist(s,t)=\tbtdistct(s,t)= \lambda \cdot\max(\varepsilon_1,\varepsilon_2)$, $\trdistdist(s,t)= \lambda \cdot 0.5$ and $\trdistdistct(s,t) = \lambda \cdot \max\{0.25 + \varepsilon_1, 0.25 + \varepsilon_2\}$.}
\end{figure}

\begin{definition}
[Tbt-trace equivalence \cite{BdNL12}]
Let $(\proc,\Act,\trans)$ be a PTS and $\mathrm{x} \in \{\mathrm{det},\mathrm{rand}\}$.
We say that $s,t \in \proc$ are in the \emph{tbt-trace preorder}, written $s \sqtbtx t$, if for each $\alpha \in \Act^{\star}$
\begin{center}
$\text{for each } \Z_s \in \resx(s) \text{ there is } \Z_t \in \resx(t) \text{ such that } \pr(\C(z_s,\alpha)) = \pr(\C(z_t,\alpha))$.
\end{center}
Then, $s,t\in \proc$ are \emph{tbt-trace equivalent}, notation $s \tbtx t$, if{f} $s \sqtbtx t$ and $t \sqtbtx s$.
\end{definition}

In \cite{BdNL12} it was proved that tbt-trace equivalences enjoy the congruence property and are full backward compatible with the fully nondeterministic and the fully probabilistic cases.

We introduce now the quantitative analogous to tbt-trace equivalences.
Processes $s,t$ are at distance $\varepsilon \ge 0$ if, \emph{for each trace} $\alpha$, for each resolution for $s$ there is a resolution for $t$ such that the two resolutions assign to $\alpha$ probabilities that \emph{differ at most by $\varepsilon$}, and vice versa.

\begin{definition}
[Tbt-trace metric]
\label{def:tbtdist}
Let $(\proc,\Act,\trans)$ be a PTS, $\lambda \in (0,1]$ and $\mathrm{x} \in \{\mathrm{det},\mathrm{rand}\}$. 
For each $\alpha \in \Act^{\star}$, the function $\tbtpredistx{\alpha,} \colon \proc \times \proc \to [0,1]$ is defined for all $s,t \in \proc$ by
\[
\tbtpredistx{\alpha,}(s,t) = \lambda^{|\alpha|-1}\; \sup_{\Z_s \in \resx(s)}\, \inf_{\Z_t \in \resx(t)} \, |\pr(\C(z_s,\alpha)) - \pr(\C(z_t,\alpha))|
\]
The \emph{tbt-trace hemimetric} and the \emph{tbt-trace metric} are the functions $\tbtpredistx{}, \tbtdistx\colon \proc \times \proc \to [0,1]$ defined for all $s,t \in \proc$ by
\begin{compactitem}
\item $\tbtpredistx{}(s,t) = \sup_{\alpha \in \Act^{\star}}\; \tbtpredistx{\alpha,}(s,t)$ 
\item $\tbtdistx(s,t) = \max \{ \tbtpredistx{}(s,t), \tbtpredistx{}(t,s)\}$.
\end{compactitem}
\end{definition}

It is not hard to see that for processes in Figure~\ref{fig:tbt-vs-td} we have $\tbtdistx(s,t) = \lambda \cdot \max(\varepsilon_1,\varepsilon_2)$ (and $s \tbtx t$ if $\varepsilon_1,\varepsilon_2=0$).
Notice that, since we consider image finite processes, we are guaranteed that for each trace $\alpha \in \Act^{\star}$ the supremum and infimum in the definition of $\tbtpredistx{\alpha,}$ are actually achieved.  
We show now that tbt-trace hemimetrics and metrics are well-defined and that their kernels are the tbt-trace preorders and equivalences, respectively.

\begin{theorem}
\label{thm:tbtdist_is_metric}
Let $(\proc,\Act,\trans)$ be a PTS, $\lambda \in (0,1]$ and $\mathrm{x} \in \{\mathrm{det},\mathrm{rand}\}$. 
Then:
\begin{enumerate}
\item The function $\tbtpredistx{}$ is a $1$-bounded hemimetric on $\proc$, with $\sqtbtx$ as kernel.
\item The function $\tbtdistx$ is a $1$-bounded pseudometric on $\proc$, with $\tbtx$ as kernel.
\end{enumerate}
\end{theorem}

\begin{example}
\label{ex:tbtdist}
Consider Figure~\ref{fig:one_for_all}.
We get $\tbtpredist{}(s_p,t)$ = $\trdistpredist(s_p,t)$ = (see Example~\ref{ex:trdist}) $\lambda \cdot \min\{p,|0.5-p|,1-p\}$.
The reason why in this particular case the two pseudometrics coincide is that each resolution for $s_p$ gives positive probability to at most one of the traces $ab$ and $ac$, so that quantifying on traces before or after quantifying on resolutions is irrelevant.

Let us evaluate now $\tbtpredist{}(t,s_p)$.
To this aim, we focus on trace $ab$ and the resolution $\Z_t$ obtained from the central $a$-branch of $t$, for which we have $\pr(\C(z_t,ab)) = 0.5$.
We need the resolution $\Z_{s_p}$ for $s_p$ that minimizes $|0.5 - \pr(\C(z_{s_p},ab))|$. 
Since for any resolutions $\Z_{s_p}$ for $s_p$ we have $\pr(\C(z_{s_p},ab)) \in \{0,p,1\}$, we infer that the resolution $\Z_{s_p}$ we are looking for satisfies $\pr(\C(z_{s_p},ab)) = p$ and, therefore, $|0.5 - \pr(\C(z_{s_p},ab))|$ = $|0.5 - p|$. 
By considering also the other resolutions for $ab$ and, then, the other traces, we can check that $\tbtpredist{}(t,s_p) = \lambda\cdot|0.5-p|$.
In Example~\ref{ex:trdist} we showed that $\trdistpredist(t,s_p) = \lambda \cdot 0.5$ for all $p \in [0,1]$.
Hence, we get $\trdistpredist(t,s_p) = \tbtpredist{}(t,s_p)$ for $p \in \{0,1\}$, and $\trdistpredist(t,s_p) > \tbtpredist{}(t,s_p)$ for $p \in (0,1)$.
This disparity is due to the fact that the trace distributions approach forced us to match the resolution for $t$ assigning positive probability to both $ab$ and $ac$, whereas in the trace-by-trace approach one never consider two traces at the same time. 
\end{example}

We conclude this section by stating that tbt-trace distances are strictly non-expansive,
As a corollary, we re-obtain the (pre)congurence properties for their kernels (proved in \cite{BdNL14}). 

\begin{theorem}
\label{thm:trace-by-trace-dist_non_exp}
All distances $\tbtpredist{}$, $\tbtpredistct{}$, $\tbtdist$, $\tbtdistct$ are strictly non-expansive.
\end{theorem}

%=======================================

\subsection{The supremal probabilities approach}
\label{sec:trace_sup}

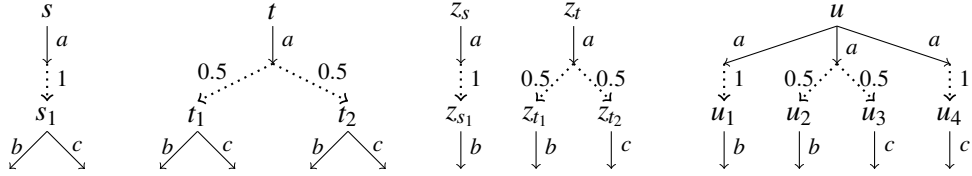
\begin{figure}[t!]
\centering
\begin{tikzpicture}
\node at (0.5,7.3){$\boldsymbol{s}$};
\draw[->](0.5,7.1)--(0.5,6.6);
\node at (0.7,6.9){\scalebox{0.8}{$a$}};
\draw[dotted, thick, ->](0.5,6.6)--(0.5,6.1);
\node at (0.7,6.4){\scalebox{0.8}{$1$}};
\node at (0.5,5.9){$\boldsymbol{s_1}$};
\draw[->](0.5,5.7)--(0,5.2);
\node at (0.1,5.5){\scalebox{0.8}{$b$}};
\draw[->](0.5,5.7)--(1,5.2);
\node at (0.9,5.5){\scalebox{0.8}{$c$}};
\node at (3.5,7.3){$\boldsymbol{t}$};
\draw[->](3.5,7.1)--(3.5,6.6);
\node at (3.7,6.9){\scalebox{0.8}{$a$}};
\draw[dotted, thick, ->](3.5,6.6)--(2.5,6.1);
\node at (2.7,6.5){\scalebox{0.8}{$0.5$}};
\draw[dotted, thick, ->](3.5,6.6)--(4.5,6.1);
\node at (4.3,6.5){\scalebox{0.8}{$0.5$}};
\node at (2.5,5.9){$\boldsymbol{t_1}$};
\draw[->](2.5,5.7)--(2,5.2);
\node at (2.1,5.5){\scalebox{0.8}{$b$}};
\draw[->](2.5,5.7)--(3,5.2);
\node at (2.9,5.5){\scalebox{0.8}{$c$}};
\node at (4.5,5.9){$\boldsymbol{t_2}$};
\draw[->](4.5,5.7)--(4,5.2);
\node at (4.1,5.5){\scalebox{0.8}{$b$}};
\draw[->](4.5,5.7)--(5,5.2);
\node at (4.9,5.5){\scalebox{0.8}{$c$}};
\node at (6,7.3){$\boldsymbol{z_s}$};
\draw[->](6,7.1)--(6,6.6);
\node at (6.2,6.9){\scalebox{0.8}{$a$}};
\draw[dotted,thick,->](6,6.6)--(6,6.1);
\node at (6.2,6.4){\scalebox{0.8}{$1$}};
\node at (6,5.9){$\boldsymbol{z_{s_1}}$};
\draw[->](6,5.7)--(6,5.2);
\node at (6.2,5.5){\scalebox{0.8}{$b$}};
\node at (7.5,7.3){$\boldsymbol{z_t}$};
\draw[->](7.5,7.1)--(7.5,6.6);
\node at (7.7,6.9){\scalebox{0.8}{$a$}};
\draw[dotted,thick,->](7.5,6.6)--(7,6.1);
\node at (7,6.4){\scalebox{0.8}{$0.5$}};
\draw[dotted,thick,->](7.5,6.6)--(8,6.1);
\node at (8,6.4){\scalebox{0.8}{$0.5$}};
\node at (7,5.9){$\boldsymbol{z_{t_1}}$};
\node at (8,5.9){$\boldsymbol{z_{t_2}}$};
\draw[->](7,5.7)--(7,5.2);
\node at (7.2,5.5){\scalebox{0.8}{$b$}};
\draw[->](8,5.7)--(8,5.2);
\node at (8.2,5.5){\scalebox{0.8}{$c$}};
\node at (11,7.3){$\boldsymbol{u}$};
\draw[->](11,7.1)--(9.5,6.6);
\node at (9.7,6.9){\scalebox{0.8}{$a$}};
\draw[->](11,7.1)--(11,6.6);
\node at (11.2,6.8){\scalebox{0.8}{$a$}};
\draw[->](11,7.1)--(12.5,6.6);
\node at (12.3,6.9){\scalebox{0.8}{$a$}};
\draw[dotted, thick, ->](9.5,6.6)--(9.5,6.1);
\node at (9.7,6.4){\scalebox{0.8}{$1$}};
\draw[dotted, thick, ->](11,6.6)--(10.5,6.1);
\node at (10.5,6.4){\scalebox{0.8}{$0.5$}};
\draw[dotted, thick, ->](11,6.6)--(11.5,6.1);
\node at (11.5,6.4){\scalebox{0.8}{$0.5$}};
\draw[dotted, thick, ->](12.5,6.6)--(12.5,6.1);
\node at (12.7,6.4){\scalebox{0.8}{$1$}};
\node at (9.5,5.9){$\boldsymbol{u_1}$};
\draw[->](9.5,5.7)--(9.5,5.2);
\node at (9.7,5.5){\scalebox{0.8}{$b$}};
\node at (10.5,5.9){$\boldsymbol{u_2}$};
\draw[->](10.5,5.7)--(10.5,5.2);
\node at (10.7,5.5){\scalebox{0.8}{$b$}};
\node at (11.5,5.9){$\boldsymbol{u_3}$};
\draw[->](11.5,5.7)--(11.5,5.2);
\node at (11.7,5.5){\scalebox{0.8}{$c$}};
\node at (12.5,5.9){$\boldsymbol{u_4}$};
\draw[->](12.5,5.7)--(12.5,5.2);
\node at (12.7,5.5){\scalebox{0.8}{$c$}};
\end{tikzpicture}
\caption{\label{fig:resolutions_distinguishing_power} Processes $s$ and $t$ are distinguished by $\tbt$, but related by $\supTr$.
We remark that $t$ and $u$ are related by all the relations in the three approaches to trace semantics.}
\end{figure}

The trace-by-trace approach improves on trace distribution approach since it supports equivalences and metrics that are compositional.
Moreover, by focusing on traces instead of resolutions, the trace-by-trace approach puts processes in Figure~\ref{fig:tbt-vs-td} in the expected relations.
However, we argue here that trace-by-trace approach on deterministic schedulers still gives some questionable results.
Take, for example, processes $s,t$ in Figure~\ref{fig:resolutions_distinguishing_power}.
We believe that these processes should be equivalent in any semantics approach, since, after performing the action $a$, they reach two distributions that should be identified, as they assign total probability $1$ to states with an identical behavior.
But, if we consider the trace $ab$, the resolution $\Z_t \in \res(t)$ in Figure~\ref{fig:resolutions_distinguishing_power} is such that $\pr(\C(z_t, ab)) = 0.5$, whereas the unique resolution for $s$ assigning positive probability to $ab$ is $\Z_s$ in Figure~\ref{fig:resolutions_distinguishing_power}, for which $\pr(\C(z_s,ab)) = 1$.
Hence no resolution in $\res(s)$ matches $\Z_t$ on trace $ab$, thus giving $\tbtdist(s.t) = \lambda \cdot 0.5$ and, consequently, $s \not\tbt t$.
This motivates to look for an alternative approach that allows us to equate processes in Figure~\ref{fig:resolutions_distinguishing_power} and, at the same time, preserves all the desirable properties of the tbt-trace semantics.

We take inspiration from the \emph{extremal probabilities} approach proposed in \cite{BdNL13}, which bases on
the comparison, for each trace $\alpha$, of both suprema and infima execution probabilities, wrt.\ resolutions, of $\alpha$: two processes are equated if they assign the same extremal probabilities to all traces.
However, reasoning on infima may cause some arguable results.
In particular, it is unclear whether such infima should be evaluated over the whole class of resolutions or over a restricted class, as for instance the resolutions in which the considered trace is actually executed.
Besides, desirable properties like the backward compatibility and compositionality are not guaranteed.
For all these reasons, we find it more reasonable to define a notion of trace equivalence, and a related metric, based on the comparison of supremal probabilities only.

Notice that, if we focus on \emph{verification}, the comparison of supremal probabilities becomes natural.
To exemplify, we let the non-probabilistic case guide us.
To verify whether a process $t$ satisfies the specification $S$, we check that whenever $S$ can execute a particular trace, then so does $t$.
Actually, only \emph{positive information} is considered: if there is a resolution for $S$ in which a given trace is executed, then this information is used to verify the equivalence.
Still, resolutions in which such a trace is not enabled are not considered.
The same principle should hold for PTSs: a process should perform all the traces enabled in $S$ and it should do it with \emph{at least} the same probability, in the perspective that the quantitative behavior expressed in the specification expresses the minimal requirements on process behavior.

Focusing on supremal probabilities means relaxing the tbt-trace approach by simply requiring that, \emph{for each trace $\alpha$} and resolution $\Z_s$ for process $s$ there is a resolution for $t$ assigning to $\alpha$ \emph{at least} the same probability given by $\Z_s$, and vice versa.

\begin{definition}
[$\bigsqcup$-trace equivalence]
\label{def:sup_trace_eq}
Let $(\proc,\Act,\trans)$ be a PTS and $\mathrm{x} \in \{\mathrm{det}, \mathrm{rand}\}$.
We say that $s,t \in \proc$ are in the \emph{$\bigsqcup$-trace preorder}, written $s \sqsupTrx t$, if for each $\alpha \in \Act^{\star}$ 
\begin{center}
$\sup_{\Z_s \in \resx(s)} \pr(\C(z_s,\alpha)) \le \sup_{\Z_t \in \resx(t)} \pr(\C(z_t,\alpha))$.
\end{center}
Then, $s,t \in \proc$ are \emph{$\bigsqcup$-trace equivalent}, notation $s \supTrx t$, if{f} $s \sqsupTrx t$ and $t \sqsupTrx s$.
\end{definition}

We stress that all good properties of trace-by-trace approach, as the backward compatibility with the fully nondeterministic and fully probabilistic cases and the non-expansiveness of the metric wrt.\ parallel composition, are preserved by the supremal probabilities approach 
(Proposition~\ref{prop:supTr_back_nondet} and Theorem~\ref{thm:tracesupdist_non_exp} below).
Let $\TrN$ denote the trace equivalence on fully nondeterministic systems \cite{BHR84} and $\TrP$ denote the one on fully-probabilistic systems \cite{JS90}.

\begin{proposition}
\label{prop:supTr_back_nondet}
Assume a PTS $P = (\proc, \Act, \trans[])$ and processes $s,t \in \proc$. Then:
\begin{enumerate}
\item If $P$ is fully-nondeterministic, then $s \supTr t \Leftrightarrow s \supTrct t \Leftrightarrow s \TrN t$.
\item If $P$ is fully-probabilistic, then $s \supTr t \Leftrightarrow s \supTrct t \Leftrightarrow s \TrP t$.
\end{enumerate}
\end{proposition}

The idea behind the quantitative analogue of $\bigsqcup$-trace equivalence is that two processes are at distance $\varepsilon \ge 0$ if, \emph{for each trace}, the supremal execution probabilities wrt.\ the resolutions of nondeterminism for the two processes \emph{dif{f}er at most by $\varepsilon$}. 

\begin{definition}
[$\bigsqcup$-trace metric]
\label{def:tracesupdist}
Let $(\proc,\Act,\trans)$ be a PTS, $\lambda \in (0,1]$ and $\mathrm{x} \in \{\mathrm{det},\mathrm{rand}\}$. 
For each $\alpha \in \Act^{\star}$, the function $\tracesuppredistx{\alpha,} \colon \proc \times \proc \to [0,1]$ is defined for all $s,t \in \proc$ by
\[
\tracesuppredistx{\alpha,}(s,t) = \max\big\{0,\lambda^{|\alpha|-1} \big(\sup_{\Z_s \in \resx(s)} \pr(\C(z_s,\alpha)) - \sup_{\Z_t \in \resx(t)} \pr(\C(z_t,\alpha))\big)\big\}.
\] 
The $\bigsqcup$-\emph{trace hemimetric} and the $\bigsqcup$-\emph{trace metric} are the functions $\tracesuppredistx{}, \tracesupdistx\colon \proc \times \proc \to [0,1]$ defined for all $s,t \in \proc$ by
\begin{compactitem}
\item $\tracesuppredistx{}(s,t) = \sup_{\alpha \in \Act^{\star}}\; \tracesuppredistx{\alpha,}(s,t)$ and
\item $\tracesupdistx(s,t) = \max \{ \tracesuppredistx{}(s,t), \tracesuppredistx{}(t,s) \}$.
\end{compactitem}
\end{definition}

We can show that $\bigsqcup$-trace hemimetrics and metrics are well-defined and that their kernels are the $\bigsqcup$-trace preorders and equivalences, respectively.

\begin{theorem}
\label{thm:tracesupdist_is_metric}
Assume a PTS $(\proc, \Act, \trans[])$, $\lambda \in (0,1]$ and $\mathrm{x} \in \{\mathrm{det},\mathrm{rand}\}$. 
Then:
\begin{enumerate}
\item The function $\tracesuppredistx{}$ is a $1$-bounded hemimetric on $\proc$, with $\sqsupTrx$ as kernel.
\item The function $\tracesupdistx$ is a $1$-bounded pseudometric on $\proc$, with $\supTrx$ as kernel.
\end{enumerate}
\end{theorem}

We conclude this section by showing that $\bigsqcup$-trace distances are strictly non-expansive. As a corollary, we infer the (pre)congruence property of their kernels.

\begin{theorem}
\label{thm:tracesupdist_non_exp}
All distances $\tracesuppredist{}$, $\tracesuppredistct{}$, $\tracesupdist$, $\tracesupdistct$  are strictly non-expansive.
\end{theorem}

\begin{remark}
We can show that the upper bounds to the distance of composed processes provided in Thms.~\ref{thm:trace-by-trace-dist_non_exp} and \ref{thm:tracesupdist_non_exp} are tight, namely for each distance $d$ considered in these theorems, there are processes $s_1,s_2,t_1,t_2$ with $d(s_1 \parallel s_2 , t_1 \parallel t_2) = d(s_1,t_1) + d(s_2,t_2) - d(s_1,t_1) \cdot d(s_2,t_2)$.
Indeed, for $z_s,z_t$ in Fig.~\ref{fig:resolutions_distinguishing_power}, with $\lambda = 1$, we have $d(z_s,t_t) = 0.5$ and $d(z_s \parallel z_s, z_t \parallel z_t) = 0.75 = 0.5 + 0.5 - 0.5 \cdot 0.5$.
\end{remark}

%=========================================

\subsection{Comparing the distinguishing power of trace metrics}

So far, we have discussed the properties of trace-based behavioral distances under different approaches.
Our aim is now to place these distances in a spectrum.
More precisely, we will order them wrt.\ their distinguishing power: given the metrics $d,d'$ on $\proc$, we write $d > d'$ if and only if $d(s,t) \ge d'(s,t)$ for all $s,t \in \proc$ and $d(u,v) > d'(u,v)$ for some $u,v \in \proc$. 

Intuitively, for trace distributions and tbt-trace semantics, the distances evaluated on deterministic schedulers are more discriminating than their randomized analogues.

\begin{theorem}
\label{thm:tr_spectrum_rand_det}
Let $(\proc,\Act,\trans)$ be a PTS, $\lambda \in (0,1]$ and $\mathrm{y} \in \{\mathrm{dis},\mathrm{tbt}\}$. 
Then:\\[0.5 ex]
\begin{inparaenum}[1.]
\item $\mathbf{h}_{\mathrm{Tr,y}}^{\lambda,\mathrm{rand}} < \mathbf{h}_{\mathrm{Tr,y}}^{\lambda,\mathrm{det}}$.
\item $\mathbf{m}_{\mathrm{Tr,y}}^{\lambda,\mathrm{rand}} < \mathbf{m}_{\mathrm{Tr,y}}^{\lambda,\mathrm{det}}$.
\end{inparaenum}
\end{theorem}

As a corollary of Theorem~\ref{thm:tr_spectrum_rand_det}, by using the relations between distances and equivalences in Theorems~\ref{thm:trdistdist_is_metric} and~\ref{thm:tbtdist_is_metric}, we re-obtain the relations $\trdist \subset \trdistct$ and $\tbt \subset \tbtct$ proved in \cite{BdNL14}. 
Moreover, also the analogous results for preorders follow. 

As one can expect, the metrics on trace distributions are more discriminating than their corresponding ones in the trace-by-trace approach.

\begin{theorem}
\label{thm:tr_spectrum_trdist_tbt}
Let $(\proc,\Act,\trans)$ be a PTS, $\lambda \in (0,1]$ and $\mathrm{x} \in \{\mathrm{det},\mathrm{rand}\}$. 
Then: \\[0.5 ex]
\begin{inparaenum}[1.]
\item $\tbtpredistx{} < \trdistpredistx$.
\item $\tbtdistx < \trdistdistx$.
\end{inparaenum}
\end{theorem}

As a corollary, by using the kernel relations given in Theorems~\ref{thm:trdistdist_is_metric} and~\ref{thm:tbtdist_is_metric}, we re-obtain the relation  $\trdistx \subset \tbtx$ proved in \cite{BdNL14} and we get $\sqdistx \subset \sqtbtx$.
Moreover, we remark that $\trdistdistct$ is not comparable with $\tbtdist$.
This is mainly due to the randomization process and it is witnessed by processes in Figure~\ref{fig:resolutions_distinguishing_power}, where $\trdistdistct(s,t)=\lambda \cdot \max\{0.25+\varepsilon_1,0.25+\varepsilon_2\}$ and $\tbtdist(s,t) = \lambda \cdot \max(\varepsilon_1,\varepsilon_2)$ and Figure~\ref{fig:tbt-vs-td}, where $\trdistdistct(s,t)=0$ and $\tbtdist(s,t) = \lambda \cdot 0.5$.

We focus now on supremal probabilities approach, that comes with a particularly interesting result: the $\bigsqcup$-trace metric on deterministic schedulers coincides with tbt-trace metrics on randomized schedulers.
Moreover, $\tracesupdist$  coincides also with its randomized version. 

\begin{theorem}
\label{thm:tr_spectrum_sup}
Assume a PTS $P = (\proc, \Act, \trans[])$ and $\lambda \in (0,1]$. 
Then:\\[0.5 ex]
\begin{inparaenum}[1.]
\item $\tracesuppredist{} = \tracesuppredistct{} = \tbtpredistct{}$.
\item $\tracesupdist = \tracesupdistct = \tbtdistct$.
\end{inparaenum}
\end{theorem}

This result is fundamental in the perspective of the application of our trace metrics to process verification: by comparing solely the suprema execution probabilities of the linear properties of interest we get same expressive power of a pairwise comparison of the probabilities in all possible randomized resolutions of nondeterminism.

Clearly, Theorem~\ref{thm:tr_spectrum_sup} together with the kernel relations from Thms~\ref{thm:tracesupdist_is_metric} and~\ref{thm:tbtdist_is_metric} imply that the relations for the supremal probabilities semantics coincide with those for the tbt-trace semantics wrt.\ randomized schedulers, ie.\ $\sqsupTr = \sqsupTrct = \sqtbtct$ and $\supTr = \supTrct = \tbtct$.

%===========================================================
% sec - testing
%==========================================================

\section{Metrics for testing}
\label{sec:test_metric}

Testing semantics~\cite{dNH84} compares processes according to their capacity to \emph{pass} a test. 
The latter is a PTS equipped with a distinguished state indicating the success of the test. 

\begin{definition}
[Test]
\label{def:test}
A \emph{nondeterministic probabilistic test transition systems} (NPT) is a finite PTS $(\OO, \Act, \trans[])$ where $\OO$ is a set of processes, called \emph{tests}, containing a distinguished \emph{success process} $\surd$ with no outgoing transitions.
We say that a computation from $o \in \OO$ is \emph{successful} if{f} its last state is $\surd$.  
\end{definition}

Given a process $s$ and a test $o$, we can consider the \emph{interaction system} among the two. 
This models the response of the process to the application of the test, so that $s$ \emph{passes} the test $o$ if there is a computation in the interaction system that reaches $\surd$.
Informally, the interaction system is the result of the parallel composition of the process with the test.

\begin{definition}
[Interaction system]
\label{def:interaction_system}
The \emph{interaction system} of a PTS $(\proc,\Act,$ $\trans[])$ and an NPT $(\OO,\Act,\trans[]_{\OO})$ is the PTS $(\proc \times \OO, \Act,\trans[]')$ where:
\begin{inparaenum}[(i)]
\item $(s,o) \in \proc \times \OO$ is called a \emph{configuration} and is \emph{successful} if{f} $o = \surd$;
\item a computation from $(s,o) \in \proc \times \OO$ is \emph{successful} if{f} its last configuration is successful.
\end{inparaenum}
\end{definition}

\noindent For $(s, o)$ and $\Z_{s,o} \in \resx(s,o)$, we let $\SC(z_{s,o})$ be the set of \emph{successful computations} from $z_{s,o}$.
For $\alpha \in \Act^{\star}$, $\SC(z_{s,o},\alpha)$ is the set of $\alpha$-compatible successful computations from $z_{s,o}$.

Testing semantics should compare processes wrt.\ their probability to pass a test.
In this Section we consider three approaches to it:
\begin{inparaenum}[(i)]
\item the \emph{may/must},
\item the \emph{trace-by-trace}, and
\item the \emph{supremal probabilities}.
\end{inparaenum}
For each approach, we present (hemi,pseudo)metrics that provide a quantitative variant of the considered testing equivalence. 
To the best of our knowledge, ours is the first attempt in this direction.

%=======================================

\subsection{The may/must approach}

In the original work on nondeterministic systems \cite{dNH84}, testing equivalence was defined via the \emph{may} and \emph{must} preorders.
The former expresses the ability of processes to pass a test.
The latter expresses the impossibility to fail a test.
When also probability is considered, these two preorders are defined, resp., in terms of \emph{suprema} and \emph{infima} success probabilities \cite{YL92}.

\begin{definition}
[May/must testing equivalence, \cite{YL92}]
Let $(\proc,\Act,\trans)$ be a PTS, $(\OO,\Act,\trans[]_O)$ an NPT and $\mathrm{x} \in \{\mathrm{det},\mathrm{rand}\}$.
We say that $s,t \in \proc$ are in the \emph{may testing preorder}, written $s \sqmayx t$, if for each $o \in \OO$
\begin{center}
$\sup_{\Z_{s,o} \in \resx_{\max}(s,o)} \pr(\SC(z_{s,o})) \le \sup_{\Z_{t,o} \in \resx_{\max}(t,o)} \pr(\SC(z_{t,o}))$.
\end{center}
Then, $s,t\in \proc$ are \emph{may testing equivalent}, written $s \mayx t$, if{f} $s \sqmayx t$ and $t \sqmayx s$.

The notions of \emph{must testing preorder}, $\sqmustx$, and \emph{must testing equivalence}, $\mustx$, are obtained by replacing the suprema in $\sqmayx$ and $\mayx$, resp., with infima.

Finally, we say that $s,t \in \proc$ are in the \emph{may/must testing preorder}, written $s \sqmaymustx t$, if $s \sqmayx t$ and $s \sqmustx t$.
They are \emph{may/must testing equivalent}, written $s \maymustx t$, if{f} $s \sqmaymustx t$ and $t \sqmaymustx s$.
\end{definition}

The quantitative analogue to may/must testing equivalence bases on the evaluation of the differences in the extremal success probabilities.
The distance between $s,t \in \proc$ is set to $\varepsilon \ge 0$ if the maximum between the difference in the suprema and infima success probabilities wrt.\ all resolutions of nondeterminism for $s$ and $t$ is at most $\varepsilon$.
We introduce a function $\omega:\OO\rightarrow (0,1]$ that assigns to each test $o$ the proper discount. 
In fact, as the success probabilities in the may/must semantics are not related to the execution of a particular trace, in general we cannot define a discount factor as we did for the trace distances.
However, a similar construction may be regained when only tests with finite depth are considered.
In that case, we could define $\omega(o) = \lambda^{\mathrm{depth}(o)}$, for $\lambda \in (0,1]$.
We will use $\mathbf{1}$ to denote the $1$ constant function.

\begin{definition}
[May/must testing metric]
Let $(\proc,\Act,\trans)$ be a PTS, $(\OO,\Act,\trans[]_O)$ an NPT, $\omega:\OO\rightarrow (0,1]$ and $\mathrm{x} \in \{\mathrm{det},\mathrm{rand}\}$. 
For each $o\in \OO$, the function $\maypredistx{o,}\colon \proc \times \proc \to [0,1]$ is defined for all $s,t \in \proc$ by
\begin{align*}
\maypredistx{o,}(s,t)=
\max\Big\{ 0 , \omega(o)\Big(\sup_{\Z_{s,o}\in \resx_{\max}(s,o)} \pr(\SC(z_{s,o})) - \sup_{\Z_{t,o}\in \resx_{\max}(t,o)} \pr(\SC(z_{t,o}))\Big) \Big\}
\end{align*}
Function $\mustpredistx{o,}\colon \proc \times \proc \to [0,1]$ is obtained by replacing the suprema in $\maypredistx{o,}$ with infima.
Given $\mathrm{y} \in \{\mathrm{may}, \mathrm{must}\}$, the $\mathrm{y}$ \emph{testing hemimetric} and the $\mathrm{y}$ \emph{testing metric} are the functions $\mathbf{h}_{\mathrm{Te,y}}^{\omega,\mathrm{x}}, \mathbf{m}_{\mathrm{Te,y}}^{\omega,\mathrm{x}} \colon \proc \times \proc \to [0,1]$ defined for all $s,t \in \proc$ by
\begin{compactitem}
\item $\mathbf{h}_{\mathrm{Te,y}}^{\omega,\mathrm{x}}(s,t) = \sup_{o \in \OO}\; \mathbf{h}_{\mathrm{Te,y}}^{o,\omega,\mathrm{x}}(s,t)$ and 
\item $\mathbf{m}_{\mathrm{Te,y}}^{\omega,\mathrm{x}}(s,t) = \max \{ \mathbf{h}_{\mathrm{Te,y}}^{\omega,\mathrm{x}}(s,t), \mathbf{h}_{\mathrm{Te,y}}^{\omega,\mathrm{x}}(t,s) \}$.
\end{compactitem}
The \emph{may/must testing hemimetric} and the \emph{may/must testing metric} are the functions $\maymustpredistx{}, \maymustdistx \colon$ $\proc \times \proc \to [0,1]$ defined for all $s,t \in \proc$ by
\begin{compactitem}
\item $\maymustpredistx{}(s,t) = \max \{ \maypredistx{}(s,t), \mustpredistx{}(s,t) \}$.
\item $\maymustdistx(s,t) = \max \{ \maydistx{}(s,t), \mustdistx{}(s,t) \}$.
\end{compactitem}
\end{definition}

\begin{theorem}
\label{thm:maymustdist_is_metric}
Let $(\proc,\Act,\trans)$ be a PTS, $\omega:\OO\rightarrow  (0,1]$, $\mathrm{x} \in \{\mathrm{det},\mathrm{rand}\}$ and $\mathrm{y} \in \{\mathrm{may},\mathrm{must}, \mathrm{mM}\}$:
\begin{enumerate}
\item The function $\mathbf{h}_{\mathrm{Te,y}}^{\lambda,\mathrm{x}}$ is a $1$-bounded hemimetric on $\proc$, with $\sqsubseteq_{\mathrm{Te,y}}^{\mathrm{x}}$ as kernel.
\item The function $\mathbf{m}_{\mathrm{Te,y}}^{\lambda,\mathrm{x}}$ is a $1$-bounded pseudometric on $\proc$, with $\sim_{\mathrm{Te,y}}^{\mathrm{x}}$ as kernel.
\end{enumerate}
\end{theorem}

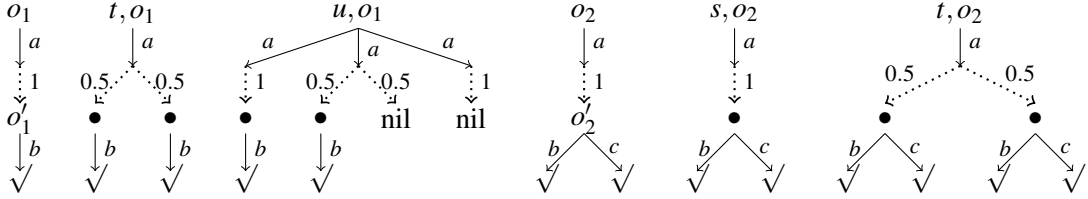
\begin{figure}[t!]
\centering
\begin{tikzpicture}
\node at (0,7.3){$\boldsymbol{o_1}$};
\draw[->](0,7.1)--(0,6.6);
\node at (0.2,6.9){\scalebox{0.8}{$a$}};
\draw[dotted, thick, ->](0,6.6)--(0,6.1);
\node at (0.2,6.4){\scalebox{0.8}{$1$}};
\node at (0,5.9){$\boldsymbol{o'_1}$};
\draw[->](0,5.7)--(0,5.2);
\node at (0.2,5.5){\scalebox{0.8}{$b$}};
\node at (0,5.1){$\surd$};
\node at (1.5,7.3){$\boldsymbol{t,o_1}$};
\draw[->](1.5,7.1)--(1.5,6.6);
\node at (1.7,6.9){\scalebox{0.8}{$a$}};
\draw[dotted, thick, ->](1.5,6.6)--(1,6.1);
\node at (1,6.4){\scalebox{0.8}{$0.5$}};
\draw[dotted, thick, ->](1.5,6.6)--(2,6.1);
\node at (2,6.4){\scalebox{0.8}{$0.5$}};
\node at (1,5.9){$\bullet$};
\draw[->](1,5.7)--(1,5.2);
\node at (1.2,5.5){\scalebox{0.8}{$b$}};
\node at (2,5.9){$\bullet$};
\draw[->](2,5.7)--(2,5.2);
\node at (2.2,5.5){\scalebox{0.8}{$b$}};
\node at (1,5.1){$\surd$};
\node at (2,5.1){$\surd$};
\node at (4.5,7.3){$\boldsymbol{u,o_1}$};
\draw[->](4.5,7.1)--(3,6.6);
\node at (3.3,6.9){\scalebox{0.8}{$a$}};
\draw[->](4.5,7.1)--(4.5,6.6);
\node at (4.7,6.8){\scalebox{0.8}{$a$}};
\draw[->](4.5,7.1)--(6,6.6);
\node at (5.7,6.9){\scalebox{0.8}{$a$}};
\draw[dotted, thick, ->](3,6.6)--(3,6.1);
\node at (3.2,6.4){\scalebox{0.8}{$1$}};
\draw[dotted, thick, ->](4.5,6.6)--(4,6.1);
\node at (4,6.4){\scalebox{0.8}{$0.5$}};
\draw[dotted, thick, ->](4.5,6.6)--(5,6.1);
\node at (5,6.4){\scalebox{0.8}{$0.5$}};
\draw[dotted, thick, ->](6,6.6)--(6,6.1);
\node at (6.2,6.4){\scalebox{0.8}{$1$}};
\node at (3,5.9){$\bullet$};
\draw[->](3,5.7)--(3,5.2);
\node at (3.2,5.5){\scalebox{0.8}{$b$}};
\node at (4,5.9){$\bullet$};
\draw[->](4,5.7)--(4,5.2);
\node at (4.2,5.5){\scalebox{0.8}{$b$}};
\node at (5,5.9){$\boldsymbol{\nil}$};
\node at (6,5.9){$\boldsymbol{\nil}$};
\node at (3,5.1){$\surd$};
\node at (4,5.1){$\surd$};
\node at (7.5,7.3){$\boldsymbol{o_2}$};
\draw[->](7.5,7.1)--(7.5,6.6);
\node at (7.7,6.9){\scalebox{0.8}{$a$}};
\draw[dotted,thick,->](7.5,6.6)--(7.5,6.1);
\node at (7.7,6.4){\scalebox{0.8}{$1$}};
\node at (7.5,5.9){$\boldsymbol{o_2'}$};
\draw[->](7.5,5.7)--(7,5.2);
\node at (7.1,5.5){\scalebox{0.8}{$b$}};
\draw[->](7.5,5.7)--(8,5.2);
\node at (7.9,5.5){\scalebox{0.8}{$c$}};
\node at (7,5.1){$\surd$};
\node at (8,5.1){$\surd$};
\node at (9.5,7.3){$\boldsymbol{s,o_2}$};
\draw[->](9.5,7.1)--(9.5,6.6);
\node at (9.7,6.9){\scalebox{0.8}{$a$}};
\draw[dotted,thick,->](9.5,6.6)--(9.5,6.1);
\node at (9.7,6.4){\scalebox{0.8}{$1$}};
\node at (9.5,5.9){$\bullet$};
\draw[->](9.5,5.7)--(9,5.2);
\node at (9.1,5.5){\scalebox{0.8}{$b$}};
\draw[->](9.5,5.7)--(10,5.2);
\node at (9.9,5.5){\scalebox{0.8}{$c$}};
\node at (9,5.1){$\surd$};
\node at (10,5.1){$\surd$};
\node at (12.5,7.3){$\boldsymbol{t,o_2}$};
\draw[->](12.5,7.1)--(12.5,6.6);
\node at (12.7,6.9){\scalebox{0.8}{$a$}};
\draw[dotted, thick, ->](12.5,6.6)--(11.5,6.1);
\node at (11.7,6.5){\scalebox{0.8}{$0.5$}};
\draw[dotted, thick, ->](12.5,6.6)--(13.5,6.1);
\node at (13.3,6.5){\scalebox{0.8}{$0.5$}};
\node at (11.5,5.9){$\bullet$};
\draw[->](11.5,5.7)--(11,5.2);
\node at (11.1,5.5){\scalebox{0.8}{$b$}};
\draw[->](11.5,5.7)--(12,5.2);
\node at (11.9,5.5){\scalebox{0.8}{$c$}};
\node at (13.5,5.9){$\bullet$};
\draw[->](13.5,5.7)--(13,5.2);
\node at (13.1,5.5){\scalebox{0.8}{$b$}};
\draw[->](13.5,5.7)--(14,5.2);
\node at (13.9,5.5){\scalebox{0.8}{$c$}};
\node at (11,5.1){$\surd$};
\node at (12,5.1){$\surd$};
\node at (13,5.1){$\surd$};
\node at (14,5.1){$\surd$};
\end{tikzpicture}
\caption{\label{fig:testing} We use the tests $o_1,o_2$ to evaluate the distance between processes $s,t,u$ in Fig.~\ref{fig:resolutions_distinguishing_power} wrt.\ testing semantics. 
$\bullet$ represents a generic configuration in the interaction system.
In all upcoming examples we will consider only the tests and traces that are significant for the evaluations of the testing metrics.}
\end{figure}

\begin{example}
\label{ex:may_must}
Consider $t,u$ in Fig~\ref{fig:resolutions_distinguishing_power} and their interactions with test $o_1$ in Fig~\ref{fig:testing}.
Clearly, $(t,o_1)$ and $(u,o_1)$ have the same suprema success probabilities.
In fact, they both have a maximal resolution assigning probability $1$ to the trace $ab$, ie.\ the only successful trace in the considered case.
As the same holds for all tests we get $\maydistx(t,u) = 0$.
Conversely, if we compare the infima success probabilities, we get $\inf_{\Z_{t,o_1} \in \resx_{\max}(t,o_1)} = 1$ since $(t,o_1)$ has only one maximal resolution corresponding to $(t,o_1)$ itself and that with probability $1$ reaches $\surd$.
Still, $\inf_{\Z_{u,o_1} \in \resx_{\max}(u,o_1)} = 0$, given by the maximal resolution corresponding to $(u,o_1) \ctrans[a] \nil$.
Hence, we can infer $\mustdistx(t,u) = \omega(o_1) \cdot |1-0| = \omega(o_1)$.
\end{example}

\noindent
We can finally observe that both $\mathbf{h}_{\mathrm{Te,y}}^{\omega,\mathrm{x}}$ and $\mathbf{m}_{\mathrm{Te,y}}^{\omega,\mathrm{x}}$ are non-expansive. 

\begin{theorem}
\label{thm:tesupinf_non_exp}
Let $\omega:\OO\rightarrow (0,1]$ and $\mathrm{y} \in \{\mathrm{may},\mathrm{must}, \mathrm{mM}\}$.
$\mathbf{h}_{\mathrm{Te,y}}^{\omega,\mathrm{x}}$ and $\mathbf{m}_{\mathrm{Te,y}}^{\omega,\mathrm{x}}$ are non-expansive.
\end{theorem}

%=============================================

\subsection{The trace-by-trace approach}

In \cite{BdNL14} it was proved that the may/must is fully backward compatible with the restricted class of processes only if the same restriction is applied to the class of tests, ie.\ if we consider resp.\ fully nondeterministic and fully probabilistic tests only.
This is due to the duplication ability of nondeterministic probabilistic tests.
However, by applying the trace-by-trace approach to testing semantics, we regain the full backward compatibility wrt.\ all tests (cf.\ \cite[Thm.~5.4]{BdNL14}).

\begin{definition}
[Tbt-testing equivalence]
Let $(\proc,\Act,\trans)$ be a PTS, $(\OO,\Act,\trans[]_O)$ an NPT, $\mathrm{x} \in \{\mathrm{det},\mathrm{rand}\}$.
We say that $s,t \in \proc$ are in the \emph{tbt-testing preorder}, written $s \sqtbtTex t$, if for each $o \in \OO$ and $\alpha \in \Act^{\star}$
\begin{center}
$\text{for each } \Z_{s,o} \in \resx_{\max}(s,o) \text{ there is } \Z_{t,o} \in \resx_{\max}(t,o) \text{ st. } \pr(\SC(z_{s,o},\alpha)) = \pr(\SC(z_{t,o},\alpha))$.
\end{center}
Then, $s,t\in \proc$ are \emph{tbt-testing equivalent}, notation $s \tbtTex t$, if{f} $s \sqtbtTex t$ and $t \sqtbtTex s$.
\end{definition}

The definition of the \emph{tbt-testing metric} naturally follows from Def.~\ref{def:tbtdist}.

\begin{definition}
[Tbt-testing metric]
Let $(\proc,\Act,\trans)$ be a PTS, $(\OO,\Act,\trans[]_O)$ an NPT, $\lambda \in (0,1]$ and $\mathrm{x} \in \{\mathrm{det},\mathrm{rand}\}$. 
For each $o \in \OO$ and $\alpha \in \Act^{\star}$, function $\testtbtpredistx{o,\alpha,} \colon \proc \times \proc \to [0,1]$ is defined for all $s,t \in \proc$ by
\[
\testtbtpredistx{o,\alpha,}(s,t) = \lambda^{|\alpha|-1}\; \sup_{\Z_{s,o} \in \resx_{\max}(s,o)}\, \inf_{\Z_{t,o} \in \resx_{\max}(t,o)} \, |\pr(\SC(z_{s,o},\alpha)) - \pr(\SC(z_{t,o},\alpha))|
\]
The \emph{tbt-testing hemimetric} and the \emph{tbt-testing metric} are the functions $\testtbtpredistx{}, \testtbtdistx\colon \proc \times \proc \to [0,1]$ defined for all $s,t \in \proc$ by
\begin{compactitem}
\item $\testtbtpredistx{}(s,t) = \sup_{o \in \OO} \; \sup_{\alpha \in \Act^{\star}}\; \testtbtpredistx{o,\alpha,}(s,t)$ 
\item $\testtbtdistx(s,t) = \max \{ \testtbtpredistx{}(s,t), \testtbtpredistx{}(t,s) \}$.
\end{compactitem}
\end{definition}

\begin{theorem}
\label{thm:testtbtdist_is_metric}
Let $(\proc,\Act,\trans)$ be a PTS, $\lambda \in (0,1]$ and $\mathrm{x} \in \{\mathrm{det},\mathrm{rand}\}$. 
Then:
\begin{enumerate}
\item The function $\testtbtpredistx{}$ is a $1$-bounded hemimetric on $\proc$, with $\sqtbtTex$ as kernel.
\item The function $\testtbtdistx$ is a $1$-bounded pseudometric on $\proc$, with $\tbtTex$ as kernel.
\end{enumerate}
\end{theorem}

\begin{example} 
Consider $s,t$ in Fig.~\ref{fig:resolutions_distinguishing_power} and their interactions with test $o_2$ in Fig.~\ref{fig:testing}.
By the same reasoning detailed in the first paragraph of Sect.~\ref{sec:trace_sup}, we get $\testtbtdist(s,t)=\lambda \cdot 0.5$ and $\testtbtdistct(s,t) = 0$.
\end{example}

When the \emph{tbt}-approach is used to define testing metrics, we get a refinement of the non-expansiveness property to strict non-expansiveness.

\begin{theorem}
\label{thm:tetbt_non_exp}
All distances $\testtbtpredist{}$, $\testtbtpredistct{}$, $\testtbtdist$, $\testtbtdistct$ are strictly non-expansive.
\end{theorem}

%=========================================

\subsection{The supremal probabilities approach}

If we focus on verification, we can use the testing semantics to verify whether a process will behave as intended by its specification in all possible environments, as modeled by the interaction with the tests.
Informally, we could see each test as a set of requests of the environment to the system: the ones ending in the success state are those that must be answered.
The interaction of the specification with the test then tells us whether the system is able to provide those answers.
Thus, an implementation has to guarantee \emph{at least} all the answers provided by the specification.
For this reason we decided to introduce also a \emph{supremal probabilities} variant of testing semantics: for each test and for each trace we compare the suprema wrt.\ all resolutions of nondeterminism of the probabilities of processes to reach success by performing the considered trace.

\begin{definition}
[$\bigsqcup$-testing equivalence]
Let $(\proc,\Act,\trans)$ be a PTS, $(\OO,\Act,\trans[]_O)$ an NPT and $\mathrm{x} \in \{\mathrm{det}, \mathrm{rand}\}$.
We say that $s,t \in \proc$ are in the \emph{$\bigsqcup$-testing preorder}, written $s \sqsupTex t$, if for each $o \in \OO$ and $\alpha \in \Act^{\star}$
\begin{center}
$\sup_{\Z_{s,o} \in \resx_{\max}(s,o)} \pr(\SC(z_{s,o},\alpha)) \le \sup_{\Z_{t,o} \in \resx_{\max}(t,o)} \pr(\SC(z_{t,o},\alpha))$.
\end{center}
Then, $s,t \in \proc$ are \emph{$\bigsqcup$-testing equivalent}, notation $s \supTex t$, if{f} $s \sqsupTex t$ and $t \sqsupTex s$.
\end{definition}

We obtain the $\bigsqcup$-\emph{testing metric} as a direct adaptation to tests of Definition~\ref{def:tracesupdist}.

\begin{definition}
[$\bigsqcup$-testing metric]
\label{def:testsupdist}
Let $(\proc,\Act,\trans)$ be a PTS, $(\OO,\Act,\trans[]_O)$ an NPT, $\lambda \in (0,1]$ and $\mathrm{x} \in \{\mathrm{det},\mathrm{rand}\}$. 
For each $o \in \OO$, $\alpha \in \Act^{\star}$, the function $\testsuppredistx{o,\alpha,} \colon \proc \times \proc \to [0,1]$ is defined for all $s,t \in \proc$ by
\[
\testsuppredistx{o,\alpha,}(s,t) = \max\Big\{0,\lambda^{|\alpha|-1} \Big(\sup_{\Z_{s,o} \in \resx_{\max}(s,o)} \!\!\!\!\!\pr(\SC(z_{s,o},\alpha)) - \sup_{\Z_{t,o} \in \resx_{\max}(t,o)}\!\!\!\!\! \pr(\SC(z_{t,o},\alpha))\Big)\Big\}.
\] 
The $\bigsqcup$-\emph{testing hemimetric} and the $\bigsqcup$-\emph{testing metric} are the functions $\testsuppredistx{}, \testsupdistx\colon \proc \times \proc \to [0,1]$ defined for all $s,t \in \proc$ by
\begin{compactitem}
\item $\testsuppredistx{}(s,t) = \sup_{o \in \OO} \; \sup_{\alpha \in \Act^{\star}}\; \testsuppredistx{o,\alpha,}(s,t)$;
\item $\testsupdistx(s,t) = \max \{ \testsuppredistx{}(s,t), \testsuppredistx{}(t,s) \}$.
\end{compactitem}
\end{definition}

\begin{theorem}
\label{thm:testsupdist_is_metric}
Let $(\proc,\Act,\trans)$ be a PTS and $\lambda \in (0,1]$ and $\mathrm{x} \in \{\mathrm{det},\mathrm{rand}\}$. 
Then:
\begin{enumerate}
\item The function $\testsuppredistx{}$ is a $1$-bounded hemimetric on $\proc$, with $\sqsupTex$ as kernel.
\item The function $\testsupdistx$ is a $1$-bounded pseudometric on $\proc$, with $\supTex$ as kernel.
\end{enumerate}
\end{theorem}

\noindent
Finally, we can show that both $\testsuppredistx{}$ and $\testsupdistx$ are strictly non-expansive.

\begin{theorem}
\label{thm:tetbtsup_non_exp}
All distances $\testsuppredist{}$, $\testsuppredistct{}$, $\testsupdist$, $\testsupdistct$ are strictly non-expansive.
\end{theorem}

\begin{remark}
For all distances $d$ considered in Thms.~\ref{thm:tesupinf_non_exp}, \ref{thm:tetbt_non_exp},  \ref{thm:tetbtsup_non_exp} and processes $z_s,z_t$ in Fig.~\ref{fig:resolutions_distinguishing_power}, with $\lambda=1$, we have $d(z_s , z_t) = 0.5$ and $d(z_s \parallel z_s , z_t \parallel z_t) = 0.75 = 0.5 + 0.5 - 0.5 \cdot 0.5$.
Hence, the upper bounds to the distance between composed processes provided in Thms.~\ref{thm:tetbt_non_exp} and \ref{thm:tetbtsup_non_exp} are tight.
We leave as a  future work the analogous result for distances considered in Thm.~\ref{thm:tesupinf_non_exp}.
\end{remark}

%=========================================

\subsection{Comparing the distinguishing power of testing metrics}

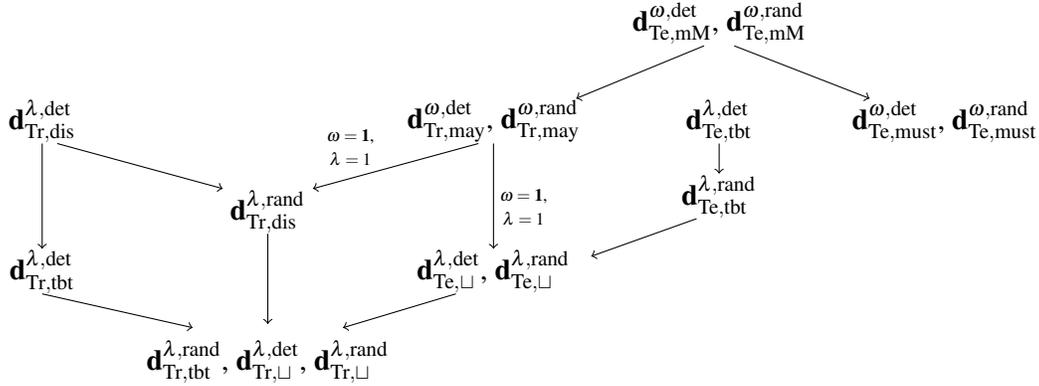
\begin{figure}
\centering
\begin{tikzpicture}
\node at (9,4.5){$\mathbf{d}_{\mathrm{Te,mM}}^{\omega,\mathrm{det}}$, $\mathbf{d}_{\mathrm{Te,mM}}^{\omega,\mathrm{rand}}$};
\draw[->](8.8,4.2)--(7.1,3.5);
\draw[->](9.2,4.2)--(11,3.5);
\node at (0,3.2){$\mathbf{d}_{\mathrm{Tr,dis}}^{\lambda,\mathrm{det}}$};
\node at (6,3.2){$\mathbf{d}_{\mathrm{Tr,may}}^{\omega,\mathrm{det}}$, $\mathbf{d}_{\mathrm{Tr,may}}^{\omega,\mathrm{rand}}$};
\node at (12,3.2){$\mathbf{d}_{\mathrm{Te,must}}^{\omega,\mathrm{det}}$, $\mathbf{d}_{\mathrm{Te,must}}^{\omega,\mathrm{rand}}$};
\draw[->](0,2.9)--(0,1.5);
\draw[->](0.2,2.9)--(2.4,2.3);
\draw[->](5.8,2.9)--(3.6,2.3);
\node at (4.1,3){\scalebox{0.6}{$\omega=\mathbf{1}$,}};
\node at (4.1,2.7){\scalebox{0.6}{$\lambda=1$}};
\draw[->](6,2.9)--(6,1.5);
\node at (6.4,2.2){\scalebox{0.6}{$\omega=\mathbf{1}$,}};
\node at (6.4,1.9){\scalebox{0.6}{$\lambda=1$}};
\node at (9,3.2){$\mathbf{d}_{\mathrm{Te,tbt}}^{\lambda,\mathrm{det}}$};
\draw[->](9,2.9)--(9,2.5);
\node at (3,2){$\mathbf{d}_{\mathrm{Tr,dis}}^{\lambda,\mathrm{rand}}$};
\draw[->](3,1.7)--(3,0.5);
\node at (9,2.2){$\mathbf{d}_{\mathrm{Te,tbt}}^{\lambda,\mathrm{rand}}$};
\draw[->](8.7,1.9)--(7.3,1.4);
\node at (0,1.2){$\mathbf{d}_{\mathrm{Tr,tbt}}^{\lambda,\mathrm{det}}$};
\draw[->](0,0.9)--(2,0.45);
\node at (6,1.2){$\mathbf{d}_{\mathrm{Te,}\sqcup}^{\lambda,\mathrm{det}}$, $\mathbf{d}_{\mathrm{Te,}\sqcup}^{\lambda,\mathrm{rand}}$};
\draw[->](5.5,0.9)--(4,0.45);
\node at (3,0){$\mathbf{d}_{\mathrm{Tr,tbt}}^{\lambda,\mathrm{rand}}$, $\mathbf{d}_{\mathrm{Tr,}\sqcup}^{\lambda,\mathrm{det}}$, $\mathbf{d}_{\mathrm{Tr,}\sqcup}^{\lambda,\mathrm{rand}}$};
\end{tikzpicture}
\caption{\label{fig:spectrum} The spectrum of trace and testing (hemi)metrics.
$d \rightarrow d'$ stands for $d > d'$.
We present only the general form with $\mathbf{d} \in \{\mathbf{h},\mathbf{m}\}$ as the relations among the hemimetrics are the same wrt.\ those among the metrics. 
The complete spectrum can be obtained by relating each metric with the respective hemimetric.}
\end{figure}

We study the distinguishing power of the testing metrics presented in this section and the trace metrics defined in Sect.~\ref{sec:trace_metric}, thus obtaining the spectrum in Fig.~\ref{fig:spectrum}. 
Firstly, we compare the expressiveness of the testing metrics wrt.\ the chosen class of schedulers.
The distinguishing power of testing metrics based on \emph{may-must} and \emph{supremal probabilities} approaches is not influenced by this choice.
Differently, in the \emph{tbt} approach, the distances evaluated on deterministic schedulers are more discriminating than their analogues on randomized schedulers. 

\begin{theorem}
\label{thm:te_spectrum_rand_det}
Let $(\proc,\Act,\trans)$ be a PTS, $\lambda \in (0,1]$, $\omega \colon \OO\to(0,1]$ $\mathrm{y} \in \{\mathrm{may},\mathrm{must}, \mathrm{mM}\}$ and $\mathbf{d} \in \{\mathbf{h},\mathbf{m}\}$:\\[0.5ex]
$\begin{array}{llll}
1.\, \mathbf{d}_{\mathrm{Te,y}}^{\omega,\mathrm{rand}} = \mathbf{d}_{\mathrm{Te,y}}^{\omega,\mathrm{det}} &
2.\, \mathbf{d}_{\mathrm{Te,tbt}}^{\lambda,\mathrm{rand}} < \mathbf{d}_{\mathrm{Te,tbt}}^{\lambda,\mathrm{det}} &
3.\, \mathbf{d}_{\mathrm{Te,}\sqcup}^{\lambda,\mathrm{rand}} = \mathbf{d}_{\mathrm{Te,}\sqcup}^{\lambda,\mathrm{det}} 
\end{array}$
\end{theorem}

From Thm.~\ref{thm:te_spectrum_rand_det}, by using the kernel relations in Thms.~\ref{thm:maymustdist_is_metric} and~\ref{thm:testtbtdist_is_metric}, we regain relations $\mayct = \may$, $\mustct = \must$, $\maymustct = \maymust$, $\tbtTe \subset \tbtTect$, and their analogues on preorders, proved in \cite{BdNL14}.
From Thm.~\ref{thm:testsupdist_is_metric} we get $\sqsupTect = \sqsupTe$ and $\supTect = \supTe$.

The strictness of the inequality in Thm.~\ref{thm:te_spectrum_rand_det}.2, is witnessed by processes $s,t$ in Fig~\ref{fig:resolutions_distinguishing_power} and their interactions with the test $o_2$ in Fig~\ref{fig:testing}.
The same reasoning applied in the first paragraph of Sect.~\ref{sec:trace_sup} to obtain $\tbtdist(s,t) = \lambda \cdot 0.5$ and $\tbtdistct(s,t) = 0$, gives $\testtbtpredist{o_2,}(t,s) = \lambda \cdot 0.5 = \testtbtdist{}(s,t)$ and $\testtbtdistct{}(s,t) = 0$.

We proceed to compare the expressiveness of each metric wrt.\ the other semantics.
Our results are fully compatible with the spectrum on probabilistic relations presented in \cite{BdNL14}.

\begin{theorem}
\label{thm:te_spectrum}
Let $(\proc,\Act,\trans)$ be a PTS, $\lambda \in (0,1]$, $\mathrm{x} \in \{\mathrm{det},\mathrm{rand}\}$ and $\mathbf{d} \in \{\mathbf{h},\mathbf{m}\}$: \\[0.5 ex]
$\begin{array}{llll}
1.\, \mathbf{d}_{\mathrm{Te,may}}^{\omega,\mathrm{x}} < \mathbf{d}_{\mathrm{Te,mM}}^{\omega,\mathrm{x}} &
2.\, \mathbf{d}_{\mathrm{Te,must}}^{\omega,\mathrm{x}} < \mathbf{d}_{\mathrm{Te,mM}}^{\omega,\mathrm{x}} &
3.\, \mathbf{d}_{\mathrm{Te,}\sqcup}^{1,\mathrm{x}} < \mathbf{d}_{\mathrm{Te,may}}^{\mathbf{1},\mathrm{x}} &
4.\, \mathbf{d}_{\mathrm{Te,}\sqcup}^{\lambda,\mathrm{x}} < \mathbf{d}_{\mathrm{Te,tbt}}^{\lambda,\mathrm{x}} \\
5.\, \mathbf{d}_{\mathrm{Tr,dis}}^{1,\mathrm{rand}} < \mathbf{d}_{\mathrm{Te,may}}^{\mathbf{1},\mathrm{x}} &
6.\, \mathbf{d}_{\mathrm{Tr,tbt}}^{\lambda,\mathrm{x}} < \mathbf{d}_{\mathrm{Te,tbt}}^{\lambda,\mathrm{x}} &
7.\, \mathbf{d}_{\mathrm{Tr,}\sqcup}^{\lambda,\mathrm{x}} < \mathbf{d}_{\mathrm{Te,}\sqcup}^{\lambda,\mathrm{x}}
\end{array}$
\end{theorem}

The following Examples prove the strictness of the inequalities in Thm.~\ref{thm:te_spectrum} and the non comparability of the (hemi)metrics as shown in Fig.~\ref{fig:spectrum}.
For simplicity, we consider only the cases of the metrics.

\begin{figure}[!t]
\centering
\begin{tikzpicture}
\node at (1.25,4.5){$\boldsymbol{s}$};
\draw[->](1.25,4.3)--(0.5,3.8);
\node at (0.7,4.2){\scalebox{0.8}{$a$}};
\draw[->](1.25,4.3)--(2,3.8);
\node at (1.8,4.2){\scalebox{0.8}{$a$}};
\draw[dotted,thick,->](0.5,3.8)--(0,3.3);
\node at (0,3.6){\scalebox{0.8}{$0.3$}};
\draw[dotted,thick,->](0.5,3.8)--(1,3.3);
\node at (0.95,3.6){\scalebox{0.8}{$0.7$}};
\draw[dotted,thick,->](2,3.8)--(1.5,3.3);
\node at (1.55,3.6){\scalebox{0.8}{$0.3$}};
\draw[dotted,thick,->](2,3.8)--(2.5,3.3);
\node at (2.5,3.6){\scalebox{0.8}{$0.7$}};
\node at (0,3.1){$\boldsymbol{s_1}$};
\node at (1,3.1){$\boldsymbol{s_2}$};
\node at (1.5,3.1){$\boldsymbol{s_3}$};
\node at (2.5,3.1){$\boldsymbol{\nil}$};
\draw[->](0,2.9)--(0,2.4);
\node at (0.2,2.7){\scalebox{0.8}{$b$}};
\draw[->](1,2.9)--(1,2.4);
\node at (1.2,2.7){\scalebox{0.8}{$c$}};
\draw[->](1.5,2.9)--(1.5,2.4);
\node at (1.7,2.7){\scalebox{0.8}{$d$}};
\node at (4.75,4.5){$\boldsymbol{t}$};
\draw[->](4.75,4.3)--(4,3.8);
\node at (4.2,4.2){\scalebox{0.8}{$a$}};
\draw[->](4.75,4.3)--(5.5,3.8);
\node at (5.3,4.2){\scalebox{0.8}{$a$}};
\draw[dotted,thick,->](4,3.8)--(3.5,3.3);
\node at (3.5,3.6){\scalebox{0.8}{$0.3$}};
\draw[dotted,thick,->](4,3.8)--(4.5,3.3);
\node at (4.45,3.65){\scalebox{0.8}{$0.7$}};
\draw[dotted,thick,->](5.5,3.8)--(5,3.3);
\node at (5.05,3.65){\scalebox{0.8}{$0.3$}};
\draw[dotted,thick,->](5.5,3.8)--(6,3.3);
\node at (6,3.6){\scalebox{0.8}{$0.7$}};
\node at (3.5,3.1){$\boldsymbol{t_1}$};
\node at (4.5,3.1){$\boldsymbol{t_2}$};
\node at (5,3.1){$\boldsymbol{t_3}$};
\node at (6,3.1){$\boldsymbol{\nil}$};
\draw[->](3.5,2.9)--(3.5,2.4);
\node at (3.7,2.7){\scalebox{0.8}{$b$}};
\draw[->](4.5,2.9)--(4.5,2.4);
\node at (4.7,2.7){\scalebox{0.8}{$d$}};
\draw[->](5,2.9)--(5,2.4);
\node at (5.2,2.7){\scalebox{0.8}{$c$}};
\node at (8.25,4.5){$\boldsymbol{s,o_2}$};
\draw[->](8.25,4.3)--(7.5,3.8);
\node at (7.7,4.2){\scalebox{0.8}{$a$}};
\draw[->](8.25,4.3)--(9,3.8);
\node at (8.8,4.2){\scalebox{0.8}{$a$}};
\draw[dotted,thick,->](7.5,3.8)--(7,3.3);
\node at (7,3.6){\scalebox{0.8}{$0.3$}};
\draw[dotted,thick,->](7.5,3.8)--(8,3.3);
\node at (7.95,3.6){\scalebox{0.8}{$0.7$}};
\draw[dotted,thick,->](9,3.8)--(8.5,3.3);
\node at (8.55,3.6){\scalebox{0.8}{$0.3$}};
\draw[dotted,thick,->](9,3.8)--(9.5,3.3);
\node at (9.5,3.6){\scalebox{0.8}{$0.7$}};
\node at (7,3.1){$\bullet$};
\node at (8,3.1){$\bullet$};
\node at (8.5,3.1){$\bullet$};
\node at (9.5,3.1){$\bullet$};
\draw[->](7,2.9)--(7,2.4);
\node at (7.2,2.7){\scalebox{0.8}{$b$}};
\draw[->](8,2.9)--(8,2.4);
\node at (8.2,2.7){\scalebox{0.8}{$c$}};
\node at (7,2.3){$\surd$};
\node at (8,2.3){$\surd$};
\node at (11.75,4.5){$\boldsymbol{t}$};
\draw[->](11.75,4.3)--(11,3.8);
\node at (11.2,4.2){\scalebox{0.8}{$a$}};
\draw[->](11.75,4.3)--(12.5,3.8);
\node at (12.3,4.2){\scalebox{0.8}{$a$}};
\draw[dotted,thick,->](11,3.8)--(10.5,3.3);
\node at (10.5,3.6){\scalebox{0.8}{$0.3$}};
\draw[dotted,thick,->](11,3.8)--(11.5,3.3);
\node at (11.45,3.65){\scalebox{0.8}{$0.7$}};
\draw[dotted,thick,->](12.5,3.8)--(12,3.3);
\node at (12.05,3.65){\scalebox{0.8}{$0.3$}};
\draw[dotted,thick,->](12.5,3.8)--(13,3.3);
\node at (13,3.6){\scalebox{0.8}{$0.7$}};
\node at (10.5,3.1){$\bullet$};
\node at (11.5,3.1){$\bullet$};
\node at (12,3.1){$\bullet$};
\node at (13,3.1){$\bullet$};
\draw[->](10.5,2.9)--(10.5,2.4);
\node at (10.7,2.7){\scalebox{0.8}{$b$}};
\draw[->](12,2.9)--(12,2.4);
\node at (12.2,2.7){\scalebox{0.8}{$c$}};
\node at (10.5,2.3){$\surd$};
\node at (12,2.3){$\surd$};
\end{tikzpicture}
\caption{\label{fig:may_vs_must} Processes $s,t$ and their interaction systems with the test $o_2$ in Fig.~\ref{fig:testing}.}
\end{figure}
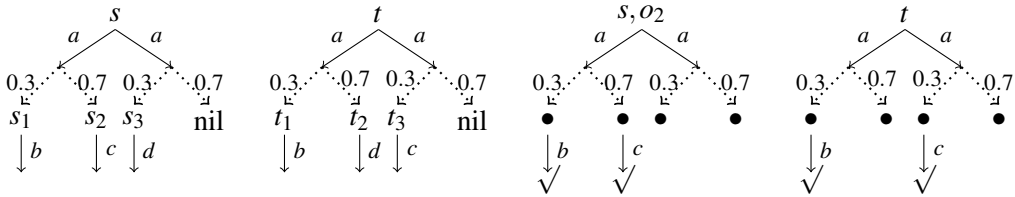

\begin{example}
\label{ex:may_vs_must}
\emph{Non comparability of $\maydistx$ with $\mustdistx$}.

In Ex.~\ref{ex:may_must} we showed that for $t,u$ in Fig.~\ref{fig:resolutions_distinguishing_power} from their interaction with the test $o_1$ in Fig.~\ref{fig:testing} we obtain that $\mustdistx(t,u) = \omega(o_1)$, whereas one can easily check that $\maydistx(t,u) = 0$.

Consider now $s,t$ and their interactions in Fig.~\ref{fig:may_vs_must} with the test $o_2$ from Fig.~\ref{fig:testing}.
Clearly, we have $\sup_{\Z_{s,o} \in \resx_{\max}(s,o)}\pr(\SC(z_{s,o})) = 1$ and $\sup_{\Z_{t,o} \in \resx_{\max}(t,o)}\pr(\SC(z_{t,o})) = 0.3$ and thus $\maydistx(s,t) = 0.7 \cdot \omega(o_2)$.
Conversely, if we consider infima success probabilities, we have $\inf_{\Z_{s,o} \in \resx_{\max}(s,o)}\pr(\SC(z_{s,o})) = 0$ and $\sup_{\Z_{t,o} \in \resx_{\max}(t,o)}\pr(\SC(z_{t,o})) = 0.3$.
Thus, $\mustdistx(s,t) = 0.3\cdot \omega(o_2)$.
\end{example}

\begin{figure}[t!]
\centering
\begin{tikzpicture}
\node at (0.5,3.7){$s$};
\draw[->](0.5,3.5)--(0.5,3);
\node at (0.7,3.3){\scalebox{0.8}{$a$}};
\draw[->,dotted,thick](0.5,3)--(0.5,2.5);
\node at (0.7,2.8){\scalebox{0.8}{$1$}};
\node at (0.5,2.3){$s_1$};
\draw[->](0.5,2.1)--(0,1.6);
\node at (0,1.9){\scalebox{0.8}{$b$}};
\draw[->](0.5,2.1)--(1,1.6);
\node at (1,1.9){\scalebox{0.8}{$b$}};
\draw[->,dotted,thick](0,1.6)--(0,1.1);
\node at (0.2,1.4){\scalebox{0.8}{$1$}};
\draw[->,dotted,thick](1,1.6)--(1,1.1);
\node at (1.2,1.4){\scalebox{0.8}{$1$}};
\node at (0,0.9){$s_2$};
\node at (1,0.9){$s_3$};
\draw[->](0,0.7)--(0,0.2);
\node at (0.2,0.5){\scalebox{0.8}{$c$}};
\draw[->](1,0.7)--(1,0.2);
\node at (1.2,0.5){\scalebox{0.8}{$d$}};
\node at (2.7,3.7){$t$};
\draw[->](2.7,3.5)--(2.2,3);
\node at (2.2,3.3){\scalebox{0.8}{$a$}};
\draw[->](2.7,3.5)--(3.2,3);
\node at (3.2,3.3){\scalebox{0.8}{$a$}};
\draw[->,dotted,thick](2.2,3)--(2.2,2.5);
\node at (2.4,2.8){\scalebox{0.8}{$1$}};
\draw[->,dotted,thick](3.2,3)--(3.2,2.5);
\node at (3.4,2.8){\scalebox{0.8}{$1$}};
\node at (2.2,2.3){$t_1$};
\node at (3.2,2.3){$t_2$};
\draw[->](2.2,2.1)--(2.2,1.6);
\node at (2.4,1.9){\scalebox{0.8}{$b$}};
\draw[->](3.2,2.1)--(3.2,1.6);
\node at (3.4,1.9){\scalebox{0.8}{$b$}};
\draw[->,dotted,thick](2.2,1.6)--(2.2,1.1);
\node at (2.4,1.4){\scalebox{0.8}{$1$}};
\draw[->,dotted,thick](3.2,1.6)--(3.2,1.1);
\node at (3.4,1.4){\scalebox{0.8}{$1$}};
\node at (2.2,0.9){$t_3$};
\node at (3.2,0.9){$t_4$};
\draw[->](2.2,0.7)--(2.2,0.2);
\node at (2.4,0.5){\scalebox{0.8}{$c$}};
\draw[->](3.2,0.7)--(3.2,0.2);
\node at (3.4,0.5){\scalebox{0.8}{$d$}};
\node at (4.9,3.7){$o^p$};
\draw[->](4.9,3.5)--(4.9,3);
\node at (5.1,3.3){\scalebox{0.8}{$a$}};
\draw[->,dotted,thick](4.9,3)--(4.4,2.5);
\node at (4.4,2.8){\scalebox{0.8}{$p$}};
\draw[->,dotted,thick](4.9,3)--(5.4,2.5);
\node at (5.4,2.8){\scalebox{0.8}{$1$-$p$}};
\node at (4.4,2.3){$o_1$};
\node at (5.4,2.3){$o_2$};
\draw[->](4.4,2.1)--(4.4,1.6);
\node at (4.6,1.9){\scalebox{0.8}{$b$}};
\draw[->](5.4,2.1)--(5.4,1.6);
\node at (5.6,1.9){\scalebox{0.8}{$b$}};
\draw[->,dotted,thick](4.4,1.6)--(4.4,1.1);
\node at (4.6,1.4){\scalebox{0.8}{$1$}};
\draw[->,dotted,thick](5.4,1.6)--(5.4,1.1);
\node at (5.6,1.4){\scalebox{0.8}{$1$}};
\node at (4.4,0.9){$o_3$};
\node at (5.4,0.9){$o_4$};
\draw[->](4.4,0.7)--(4.4,0.2);
\node at (4.6,0.5){\scalebox{0.8}{$c$}};
\draw[->](5.4,0.7)--(5.4,0.2);
\node at (5.6,0.5){\scalebox{0.8}{$d$}};
\node at (4.4,0.1){$\surd$};
\node at (5.4,0.1){$\surd$};
\node at (8.1,3.7){$s,o^p$};
\draw[->](8.1,3.5)--(8.1,3);
\node at (8.3,3.3){\scalebox{0.8}{$a$}};
\draw[->,dotted,thick](8.1,3)--(7.1,2.5);
\node at (7.1,2.8){\scalebox{0.8}{$p$}};
\draw[->,dotted,thick](8.1,3)--(9.1,2.5);
\node at (9.1,2.8){\scalebox{0.8}{$1$-$p$}};
\node at (7.1,2.3){$\bullet$};
\node at (9.1,2.3){$\bullet$};
\draw[->](7.1,2.1)--(6.6,1.6);
\node at (6.6,1.9){\scalebox{0.8}{$b$}};
\draw[->](7.1,2.1)--(7.6,1.6);
\node at (7.6,1.9){\scalebox{0.8}{$b$}};
\draw[->](9.1,2.1)--(8.6,1.6);
\node at (8.6,1.9){\scalebox{0.8}{$b$}};
\draw[->](9.1,2.1)--(9.6,1.6);
\node at (9.6,1.9){\scalebox{0.8}{$b$}};
\draw[->,dotted,thick](6.6,1.6)--(6.6,1.1);
\node at (6.8,1.4){\scalebox{0.8}{$1$}};
\draw[->,dotted,thick](9.6,1.6)--(9.6,1.1);
\node at (9.8,1.4){\scalebox{0.8}{$1$}};
\node at (6.6,0.9){$\bullet$};
\node at (9.6,0.9){$\bullet$};
\draw[->](6.6,0.7)--(6.6,0.2);
\node at (6.8,0.5){\scalebox{0.8}{$c$}};
\draw[->](9.6,0.7)--(9.6,0.2);
\node at (9.8,0.5){\scalebox{0.8}{$d$}};
\node at (6.6,0.1){$\surd$};
\node at (9.6,0.1){$\surd$};
\node at (12.3,3.7){$t,o^p$};
\draw[->](12.3,3.5)--(11.3,3);
\node at (11.3,3.3){\scalebox{0.8}{$a$}};
\draw[->,dotted,thick](11.3,3)--(10.8,2.5);
\node at (10.8,2.8){\scalebox{0.8}{$p$}};
\draw[->,dotted,thick](11.3,3)--(11.8,2.5);
\node at (11.8,2.8){\scalebox{0.8}{$1$-$p$}};
\draw[->](12.3,3.5)--(13.3,3);
\node at (13.3,3.3){\scalebox{0.8}{$a$}};
\draw[->,dotted,thick](13.3,3)--(12.8,2.5);
\node at (12.8,2.8){\scalebox{0.8}{$p$}};
\draw[->,dotted,thick](13.3,3)--(13.8,2.5);
\node at (13.8,2.8){\scalebox{0.8}{$1$-$p$}};
\node at (10.8,2.3){$\bullet$};
\node at (11.8,2.3){$\bullet$};
\node at (12.8,2.3){$\bullet$};
\node at (13.8,2.3){$\bullet$};
\draw[->](10.8,2.1)--(10.8,1.6);
\node at (11,1.9){\scalebox{0.8}{$b$}};
\draw[->](11.8,2.1)--(11.8,1.6);
\node at (12,1.9){\scalebox{0.8}{$b$}};
\draw[->](12.8,2.1)--(12.8,1.6);
\node at (13,1.9){\scalebox{0.8}{$b$}};
\draw[->](13.8,2.1)--(13.8,1.6);
\node at (14,1.9){\scalebox{0.8}{$b$}};
\draw[->,dotted,thick](10.8,1.6)--(10.8,1.1);
\node at (11,1.4){\scalebox{0.8}{$1$}};
\draw[->,dotted,thick](13.8,1.6)--(13.8,1.1);
\node at (14,1.4){\scalebox{0.8}{$1$}};
\node at (10.8,0.9){$\bullet$};
\node at (13.8,0.9){$\bullet$};
\draw[->](10.8,0.7)--(10.8,0.2);
\node at (11,0.5){\scalebox{0.8}{$c$}};
\draw[->](13.8,0.7)--(13.8,0.2);
\node at (14,0.5){\scalebox{0.8}{$d$}};
\node at (10.8,0.1){$\surd$};
\node at (13.8,0.1){$\surd$};
\end{tikzpicture}
\caption{\label{fig:isolation} Processes $s,t$ are such that $\mathbf{d}_{\mathrm{Te,tbt}}^{1,\mathrm{x}}(s,t) = 0$ and $\mathbf{d}_{\mathrm{Te,must}}^{\mathbf{1},\mathrm{x}}(s,t) = 0.5$, as witnessed by the test $o^{1/2}$.}
\end{figure}
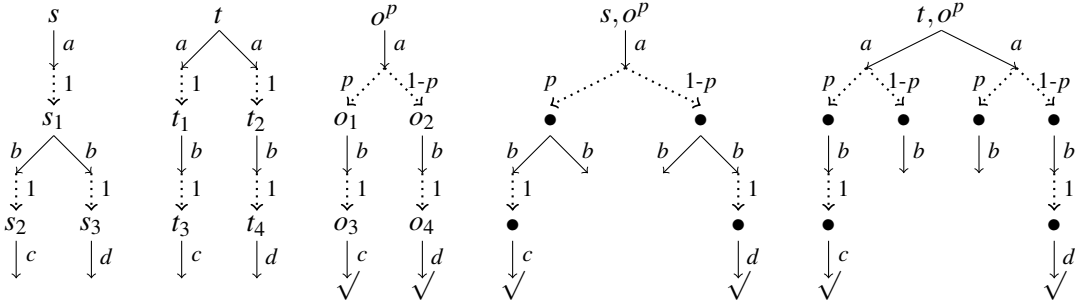

\begin{example}
\label{ex:must_vs_all}
\emph{Non comparability of $\mathbf{m}_{\mathrm{Te,must}}^{\mathbf{1},\mathrm{x}}$ with $\mathbf{m}_{\mathrm{Te,}\sqcup}^{1,\mathrm{x}}$, $\mathbf{m}_{\mathrm{Te,tbt}}^{1,\mathrm{x}}$, $\mathbf{m}_{\mathrm{Tr,dis}}^{1,\mathrm{x}}$, $\mathbf{m}_{\mathrm{Tr,tbt}}^{1,\mathrm{x}}$ and $\mathbf{m}_{\mathrm{Tr,}\sqcup}^{1,\mathrm{x}}$}.

We start with $\mathbf{m}_{\mathrm{Te,}\sqcup}^{1,\mathrm{x}}$.
Form Ex.~\ref{ex:may_must} we know that for $t,u$ in Fig.~\ref{fig:resolutions_distinguishing_power} it holds $\mathbf{m}_{\mathrm{Te,must}}^{\mathbf{1},\mathrm{x}}(t,u) = \mathbf{1}$.
Since both $t$ and $u$ have maximal resolutions giving probability $1$ to either $ab$ or $ac$, we get $\mathbf{m}_{\mathrm{Te,}\sqcup}^{1,\mathrm{x}}(t,u) =0$.
Consider now $s,t$ in Fig.~\ref{fig:may_vs_must}.
In Ex.~\ref{ex:may_vs_must} we showed that $\mathbf{m}_{\mathrm{Te,must}}^{\mathbf{1},\mathrm{x}}(s,t) = 0.3$.
From the interaction systems in Fig.~\ref{fig:may_vs_must}, by considering the superma success probabilities of trace $ac$, we obtain that $\mathbf{m}_{\mathrm{Te,}\sqcup}^{1,\mathrm{x}} = 0.4$.

Next we deal with the tbt-testing metrics.
Consider $s,t$ in Fig.~\ref{fig:isolation} and the family of tests $O = \{o^p \mid p \in (0,1)\}$, each duplicating the actions $b$ in the interaction with $s$ and $t$.
For each $o^p \in O$, $\inf_{\Z_{s,o^p} \in \resx_{\max}(s,o^p)} \pr(\SC(z_{s,o^p})) = 0$ and $\inf_{\Z_{t,o^p} \in \resx_{\max}(t,o^p)} \pr(\SC(z_{t,o^p})) = \min\{p,1-p\}$, thus giving $\mathbf{h}_{\mathrm{Te,must}}^{o^p,\mathbf{1},\mathrm{x}}(t,s) = \min\{p,1-p\}$. 
One can then easily check that $\mathbf{m}_{\mathrm{Te,must}}^{\mathbf{1},\mathrm{x}}(s,t) = \sup_{p \in (0,1)}\min\{p,1-p\} = 0.5$.
Conversely, as the tbt-testing metric compares the success probabilities related to the execution of a single trace per time, we get $\mathbf{m}_{\mathrm{Te,tbt}}^{1,\mathrm{x}}(s,t) = 0$.
Notice that in the case of randomized schedulers, all the randomized resolutions for $t,o^p$ combining the two $a$-moves can be matched by $s,o^p$ by combining the $b$-moves and vice versa.
Consider now $s,t$ in Fig.~\ref{fig:may_vs_must}.
Even under randomized schedulers, the tbt-testing distance on them is given by the difference in the success probability of the trace $ac$ (or equivalently $ad$) and thus $\mathbf{m}_{\mathrm{Te,tbt}}^{1,\mathrm{x}}(s,t) = 0.4$.
However, we have already showed that $\mathbf{m}_{\mathrm{Te,must}}^{\mathbf{1},\mathrm{x}}(s,t) = 0.3$.

Finally, we consider the case of trace distances.
Consider $t,u$ in Fig.~\ref{fig:resolutions_distinguishing_power}.
Clearly, $\mathbf{m}_{\mathrm{Tr,dis}}^{1,\mathrm{x}}(t,u) = \mathbf{m}_{\mathrm{Tr,tbt}}^{1,\mathrm{x}}(t,u) = \mathbf{m}_{\mathrm{Tr,}\sqcup}^{1,\mathrm{x}}(t,u) = 0$.
However, in Ex.~\ref{ex:may_must} we showed that $\mathbf{m}_{\mathrm{Te,must}}^{\mathbf{1},\mathrm{x}}(t,u) = 1$.
Consider now $s,t$ in Fig.~\ref{fig:may_vs_must}.
We have that $\mathbf{m}_{\mathrm{Te,must}}^{\mathbf{1},\mathrm{x}}(s,t) = 0.3$, but $\mathbf{m}_{\mathrm{Tr,dis}}^{1,\mathrm{x}}(s,t) = 0.7$ and $\mathbf{m}_{\mathrm{Tr,tbt}}^{1,\mathrm{x}}(s,t) = \mathbf{m}_{\mathrm{Tr,}\sqcup}^{1,\mathrm{x}}(s,t) = 0.4$.
\end{example}

\begin{example}
\label{ex:may_vs_all}
\emph{Non comparability of $\mathbf{m}_{\mathrm{Te,may}}^{\mathbf{1},\mathrm{x}}$ with $\mathbf{m}_{\mathrm{Te,tbt}}^{1,\mathrm{x}}$, $\mathbf{m}_{\mathrm{Tr,dis}}^{1,\mathrm{det}}$ and $\mathbf{m}_{\mathrm{Tr,tbt}}^{1,\mathrm{det}}$}.

For the tbt-testing metrics, consider $s,t$ in Fig.~\ref{fig:isolation}.
In Ex.~\ref{ex:must_vs_all} we showed that $\mathbf{m}_{\mathrm{Te,tbt}}^{1,\mathrm{x}}(s,t) = 0$.
However, the same reasoning giving $\mathbf{m}_{\mathrm{Te,must}}^{\mathbf{1},\mathrm{x}}(s,t) = 0.5$, can be applied on suprema success probabilities thus giving $\mathbf{m}_{\mathrm{Te,may}}^{\mathbf{1},\mathrm{x}}(s,t) = 0.5$.
Consider now $t,u$ in Fig.~\ref{fig:resolutions_distinguishing_power} and their interactions with test $o_1$ in Fig.~\ref{fig:testing}.
As we consider maximal resolutions only, for both classes of schedulers, the success probability of trace $ab$ evaluates to $1$ on $t,o_1$, whereas on $u,o_1$ it evaluates to $0$, due to the maximal resolution corresponding to the rightmost $a$-branch.
Hence $\mathbf{m}_{\mathrm{Te,tbt}}^{1,\mathrm{x}}(t,u)  = \lambda$, whereas one can easily check that $\mathbf{m}_{\mathrm{Te,may}}^{\mathbf{1},\mathrm{x}}(t,u)  = 0$.

We now proceed to the case of trace distances.
For $s,t$ in Fig.~\ref{fig:isolation}, we showed that $\mathbf{m}_{\mathrm{Te,may}}^{\mathbf{1},\mathrm{x}}(s,t)  = 0.5$.
However, as both processes have a single resolution each allowing them to execute either trace $abc$ or $abd$, we can infer that $\mathbf{m}_{\mathrm{Tr,dis}}^{1,\mathrm{x}}(s,t) = \mathbf{m}_{\mathrm{Tr,tbt}}^{1,\mathrm{x}}(s,t) = 0$. 
Notice, that this also shows the strictness of the relation $\mathbf{m}_{\mathrm{Tr,dis}}^{1,\mathrm{rand}} < \mathbf{m}_{\mathrm{Te,may}}^{\mathbf{1},\mathrm{x}}$.
Consider now $s,t$ in Fig.~\ref{fig:resolutions_distinguishing_power}.
As discussed in Sect.~\ref{sec:trace_sup} we have that $\mathbf{m}_{\mathrm{Tr,dis}}^{1,\mathrm{det}}  \ge \mathbf{m}_{\mathrm{Tr,tbt}}^{1,\mathrm{det}}(s,t) = 0.5$.
However, one can easily check that $\mathbf{m}_{\mathrm{Te,may}}^{\mathbf{1},\mathrm{x}}(s,t) = 0$.
\end{example}

\begin{example}
\label{ex:may_sup}
\emph{Strictness of $\mathbf{m}_{\mathrm{Te,}\sqcup}^{1,\mathrm{x}} < \mathbf{m}_{\mathrm{Te,may}}^{\mathbf{1},\mathrm{x}}$}.

Consider $s,t$ in Fig.~\ref{fig:may_vs_must}.
In Ex.~\ref{ex:may_vs_must} we have shown that $\mathbf{m}_{\mathrm{Te,may}}^{\mathbf{1},\mathrm{x}}(s,t) = 0.7$.
However, since the supremal probability approach to testing proceeds in a trace-by-trace fashion, the $\sqcup$-testing distance  is given by the difference in the success probability of the trace $ac$ (or $ad$) and thus $\mathbf{m}_{\mathrm{Te,}\sqcup}^{1,\mathrm{x}}(s,t) = 0.4$. 
\end{example}

\begin{example}
\label{ex:tbt_sup}
\emph{Strictness of $\testsupdistx < \testtbtdistx$}.

We stress that this relation is due to the restriction to maximal resolutions, necessary to reason on testing semantics.
Consider now $t,u$ in Fig.~\ref{fig:resolutions_distinguishing_power} and their interactions with test $o_1$ in Fig.\ref{fig:testing}.
In Ex.\ref{ex:may_vs_all} we have shown that $\testtbtdistx(t,u) = \lambda$.
However, one can easily check that $\testsupdistx(t,u) = 0$.
\end{example}

\begin{example}
\label{ex:test_trace}
\emph{Strictness of $\tbtdistx < \testtbtdistx$ and of $\tracesupdistx < \testsupdistx$}.

For $\mathbf{m}_{\mathrm{Tr,tbt}}^{\lambda,\mathrm{x}} < \mathbf{m}_{\mathrm{Te,tbt}}^{\lambda,\mathrm{x}}$ consider $t,u$ in Fig.~\ref{fig:resolutions_distinguishing_power} and the test $o_1$ in Fig.~\ref{fig:testing}, by which we get $\tbtdistx(t,u) = 0$ and $\testtbtdistx(t,u) = \lambda$. 
Similarly, for $\mathbf{m}_{\mathrm{Tr,}\sqcup}^{\lambda,\mathrm{x}} < \mathbf{m}_{\mathrm{Te,}\sqcup}^{\lambda,\mathrm{x}}$ consider $s,t$ in Fig.~\ref{fig:tbt-vs-td} with $\varepsilon_1=\varepsilon_2=0$.
We have $\tracesupdistx(s,t) = 0$ and $\testsupdistx(s,t) = \lambda \cdot 0.5$, given by the test $o$ corresponding to the leftmost branch of $s$.
\end{example}

%================================
% sec - related and conclusion
%==============================

\section{Related and future work}

Trace metrics have been thoroughly studied on quantitative systems, as testified by the spectrum of distances, defined as the generalization of a chosen trace distance, in \cite{FL14} and the one on Metric Transition Systems (MTSs) in \cite{AFS09}.
The great variety in these models and the PTSs prevent us to compare the obtained results in detail.
Notably, in \cite{AFS09} the trace distance is based on a \emph{propositional distance} defined over \emph{valuations} of atomic propositions that characterize the MTS.
If on one side such valuation could play the role of the probability distributions in the PTS, it is unclear whether we could combine the ground distance on atomic propositions and the propositional distance, to obtain trace distances comparable to ours.
In \cite{BBLM15,DHKP16} trace metrics on Markov Chains (MCs) are defined as total variation distances on the cones generated by traces.
As in MCs probability depends only on the current state and not on nondeterminism, our quantification over resolutions would be trivial on MCs, giving a total variation distance. 

Although ours is the first proposal of a metric expressing testing semantics, testing equivalences for probabilistic processes have been studied also in \cite{DvGHM08,BdNL12,BdNL13}.
In detail, \cite{DvGHM08} proposed notions of probabilistic \emph{may/must testing} for a Kleisli lifting of the PTS model, ie.\ the transition relation is lifted to a relation $(\rightarrow)^{\dagger} \subseteq (\ProbDist{\proc} \times \Act \times \ProbDist{\proc})$ taking distributions over processes to distributions over processes.
Again, the disparity in the two models prevents us from thoroughly comparing the proposed testing relations.

As future work, we aim to extend the spectrum of metrics to (bi)simulation metrics \cite{DGJP04} and to metrics on different semantic models, and to study their logical characterizations and compositional properties on the same line of \cite{CGT16a,CGT16b,CGT18}.
Further, we aim to provide efficient algorithms for the evaluation of the proposed metrics and to develop a tool for quantitative process verification: we will use the distance between a process and its specification to quantify how much that process satisfies a given property. 

\paragraph*{Acknowledgements}
I wish to thank Michele Loreti and Simone Tini for fruitful discussions, and the anonymous referees for their valuable comments and suggestions that helped to improve the paper.

%===========================================

\bibliographystyle{eptcs}
\bibliography{Bibliography}

%\newpage
%\appendix
%\input{express18_Appendix}

\end{document}